\documentclass[useAMS,usenatbib]{mn2e}
\usepackage{graphicx}
\usepackage{txfonts}
\usepackage{natbib}
\usepackage{color}
\newcommand{\xte}{{\it RXTE}}

\def\gsim{\mathrel{\hbox{\rlap{\hbox{\lower4pt\hbox{$\sim$}}}\hbox{$>$}}}}
\def\lsim{\mathrel{\hbox{\rlap{\hbox{\lower4pt\hbox{$\sim$}}}\hbox{$<$}}}}

\begin{document}
   \title[]{Power-Colours: Simple X-ray Binary Variability Comparison}

   \author[L. M. Heil, P. Uttley and M. Klein-Wolt]{L. M. Heil$^1$, P. Uttley$^1$ and M. Klein-Wolt$^{2,3}$\\
$^{1}$ Anton Pannekoek Institute, University of Amsterdam, Science Park 904, 1098 XH Amsterdam, The Netherlands \\
$^{2}$Department of Astrophysics, Research Institute of Mathematics, Astrophysics and Particle Physics, Radboud University Nijmegen,\\
Heijendaalseweg 135, 6525 AJ Nijmegen, The Netherlands\\
$^{3}$Science \& Technology, Olof Palmestraat 14, 2616 LR Delft, The Netherlands}

   \date{Draft \today}

   \pagerange{\pageref{firstpage}--\pageref{lastpage}} \pubyear{2002}

   \maketitle
   
   \label{firstpage}

   \begin{abstract}
	We demonstrate a new method of variability classification using observations of black hole X-ray binaries. Using `power colours' -- ratios of integrated power in different Fourier frequency bands -- we can clearly differentiate different canonical black hole states as the objects evolve during outburst. We analyse ($\sim$ 2400) {\it Rossi X-ray Timing Explorer} observations of 12 transient low mass black hole X-ray binaries and find that the path taken around the power colour-colour diagram as the sources evolve is highly consistent from object to object. We discuss how the consistency observed in the power colour-colour diagram between different objects allows for easy state classification based on only a few observations, and show how the power-spectral shapes can be simply classified using a single parameter, the power-spectral `hue'.  To illustrate the benefits of our simple model-independent approach, we show that the persistent high mass X-ray binary Cyg~X-1 shows very similar power-spectral evolution to the transient black hole sources, with the main difference being caused by a combination of a lack of quasi-periodic oscillations and an excess of low-frequency power-law noise in the Cyg X-1 power spectra during the transitional state.  We also compare the transient objects to the neutron star atoll source Aquila X-1, demonstrating that it traces a different path in the power colour-colour plot.  Thus, power-colours could be an effective method to classify newly discovered X-ray binaries.  
\end{abstract}

 

\section{Introduction}
\label{sect:intro}
The X-ray emission from black hole X-ray binaries shows distinct changes over time, interpreted as evidence that the structure of the accretion flow and the X-ray emitting regions close to the black hole evolve over the course of the outburst. For the transient, black hole low mass X-ray binaries (LMXBs) this behaviour is classified into a series of common ``states'', according to both the energy spectral and timing properties averaged over an observation \citep[see reviews by][]{Remillard06, Belloni10}.  LMXBs rise in flux out of quiescence in a hard state with the energy spectra dominated by strong power law emission. This emission is interpreted as originating from Compton scattering within a hot, optically thin component close to the compact object \citep[see e.g][]{Done07}. In the lightcurve, variability is observed over a wide range of different timescales with sharp quasi-periodic features representing strong variations over a narrow range of frequencies and broader, less coherent components \citep[see e.g][]{Wijnands99, KleinWolt08}. 

Both the energy spectra and the power spectra show evolution within the hard state as the X-ray flux from the source increases. The energy spectra become softer and emission from a blackbody-like component, associated with an optically thick accretion disc, becomes stronger. In addition the peak timescales for the power spectral features shorten and the variability becomes concentrated in a narrower frequency range (0.1-10 Hz). This is typically close to the peak flux of the outburst and is classified as the hard-intermediate state (HIMS). The reduction in timescales has been interpreted as evidence for the inner edge of the disc moving closer to the black hole. As the disc inner edge moves closer, increased disc photon flux incident on the hot corona causes cooling thus making the power-law steeper \citep[see e.g.][]{Done07}.  At some point the broad-band variability disappears leaving dominant quasi-periodic features, often with a fundamental component at $\sim$ 6 Hz and some weak variability on short timescales -- this marks the transition into a soft intermediate state (SIMS) \citep[see e.g.][]{Belloni10b}. Little evidence for distinct differences is found in the energy spectra between the HIMS and the SIMS \citep{Dunn10}, even though these dramatic changes in the broadband variability are observed. After this point the energy spectra are typically observed to be dominated by the disc blackbody component and the level of variability is very low ($\lsim$ 3 per cent fractional rms). The source flux decays whilst the object is in the soft state and the object eventually transitions back into a hard state before returning to quiescence.  The spectral evolution of BH LMXBs over time is typically illustrated using a hardness-intensity diagram \citep{Homan01, Belloni04}. In this simple plot, the ratio of hard to soft X-ray flux when compared to the total emission follows a  hysteresis curve as a LMXB goes through an outburst (see e.g. figure 1 in \cite{KleinWolt08}).  

Given the clear evolution of variability properties through the different states, it is interesting to see how timing-based classification can be used, not only to identify states but also to help identify the physical changes linked to the spectral and timing evolution.  Recently, \cite{Munoz-Darias11} demonstrated that it is possible to describe the evolution of the power spectra using an rms-intensity diagram \citep{Munoz-Darias11, Heil12}, where the rms is calculated as the square-root of the integrated area underneath the power spectra. Although the rms-intensity plot clearly demonstrates the evolution of the total power, it does not contain information about the shape of the power spectrum which makes direct comparison between different states in objects difficult, since changes in rms due to changes in power-spectral shape cannot be disentangled from changes due to power-spectral normalisation. 

An alternative to simple rms variability measurement is to fit the power spectrum with an empirical model, typically with a series of Lorentzians of different widths or bending or broken power laws \citep[see e.g.][]{Nowak00}. These models do not have a strict physical model underpinning them \citep[although see][]{Ingram09} but they describe the power spectral shape fairly well. Common correlations are observed between the peak frequencies of the Lorentzians as the source evolves through different states \citep{Wijnands99}. In the hard and hard-intermediate states all peak frequencies for the components are observed to increase together as the flux increases \citep[see e.g.][]{Axelsson05, KleinWolt08}. However, there are some difficulties in fitting models which are not underpinned by physical descriptors. Firstly, The definition of characteristic frequencies is clearly strongly dependent on the model chosen. Secondly, fitting is complicated in observations when the signal to noise ratio is low as it can be difficult to discern which features are present. Finally, whole-scale global changes to the power spectra, such as those commonly observed in transitions around the intermediate states make identification of individual components difficult to track and identify. All of these features complicate attempts to make direct comparisons between different objects and states. 

Therefore, to address the complications with current methods of classifying timing behaviour, and inspired by the use of energy spectral `colours' or hardness ratios we develop in this paper a simple `colour based' approach to classifying power spectra, by defining `power colours': ratios of integrated power (i.e. variance) measured over different Fourier frequency ranges. We apply this approach to the huge archive of observations of transient black hole LMXBs obtained by the {\it Rossi X-ray Timing Explorer} ({\it RXTE}).  Power-colours can be measured as long as the rms-variance can be measured in all set frequency bands, meaning that it is effective even in observations with low signal to noise (i.e. even when the power is very low but just above the Poisson noise).  We find that the different objects follow a distinct and consistent track in a power-`colour-colour' diagram  and show that this track lends itself to a single-value parameterisation of the power-spectral shape (and thus source state), which we call the `hue' of the power spectrum.  Finally we demonstrate that the method allows for direct comparison of the behaviour of different classes of source, namely the LMXBs and the high mass black hole X-ray binary Cyg X-1, as well as the accreting neutron star atoll source Aquila X-1, revealing intriguing similarities and differences with the black hole LMXBs.

\begin{figure*}
\begin{center}

	\includegraphics[width=13 cm, angle=90]{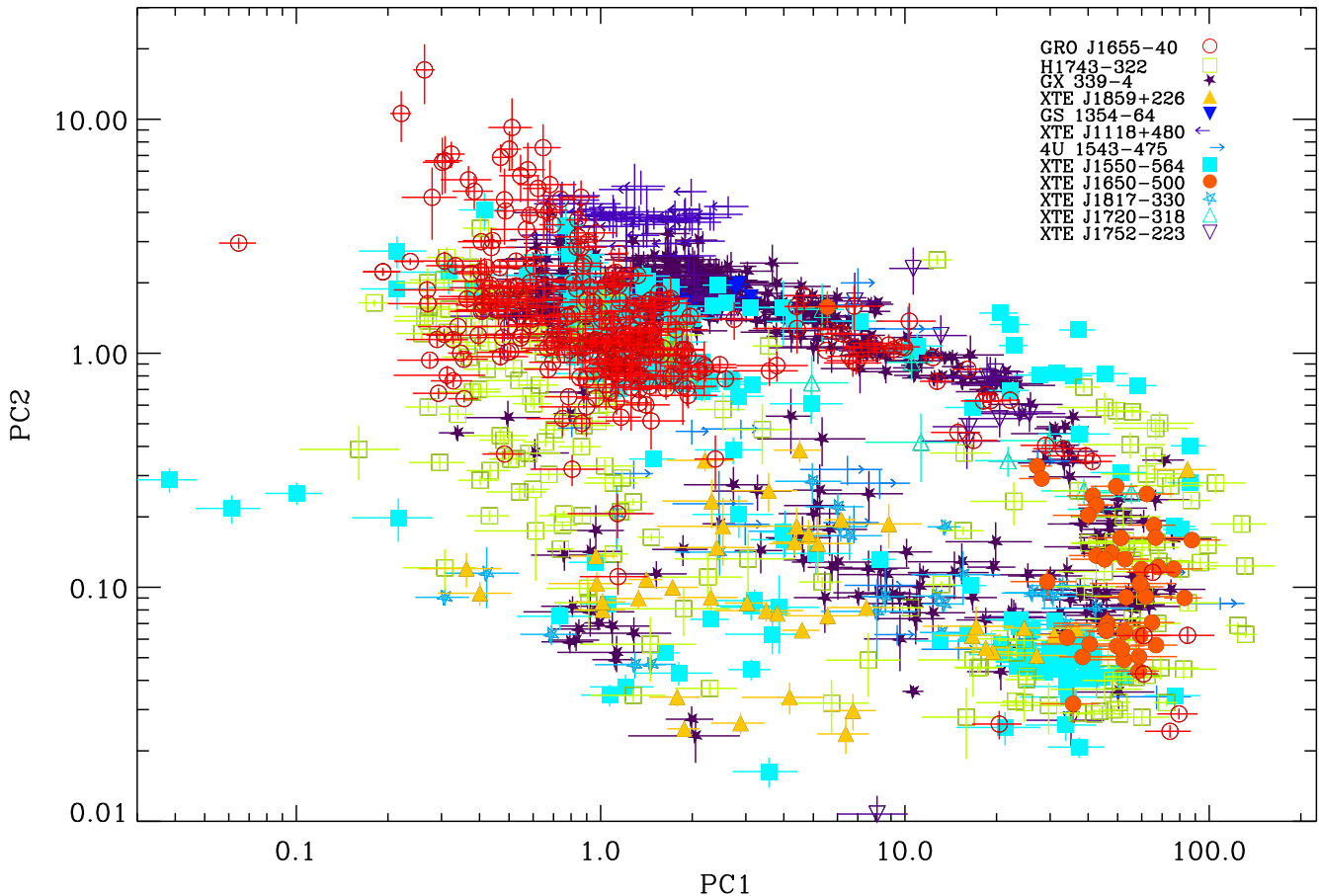}
\end{center}
\caption{Power-diagram for the ratios shown for all observations of the transient black holes included within our sample. All objects are individually colour-coded according to the legend. PC1 = variance in 0.25-2.0 Hz / 0.0039-0.031 Hz and PC2 = variance in 0.031-0.25 Hz / 2.0-16.0 Hz}
\label{fig:srccols}
\end{figure*}

\begin{table}
\centering
\begin{tabular}{lrr}
\hline\hline
Source Name &  Total Obs. & Good Obs.  \\
(1)      & (2) & (3)  \\

\hline
GX 339-4 & 619 & 457  \\
XTE J1118+480 & 65 & 48 \\
GS 1354-64 & 7 & 7 \\ 
4U 1543-475 & 51 & 21 \\
XTE J1550-564 & 268 & 201 \\
XTE J1650-500 & 97 & 41 \\
GRO J1655-40 & 438 & 312 \\
XTE J1720-318 & 85 & 10 \\
H 1743-322 & 451 & 179 \\
XTE J1752-223 & 143 & 64 \\
XTE J1817-330 & 129 & 26 \\
XTE J1859+226 & 107 & 44 \\
\hline
Cygnus X-1 & 1465 & 1414 \\
\hline
Aquila X-1 & 529 & 238 \\
\hline
Total & 4454 & 3062 \\
\hline
\multicolumn{3}{c}{{\it }}\\
\end{tabular}
\caption{(1) Source Name; (2) Total number of observations going into the sample; (3) ``Good'' observations, defined as those where the variance is constrained as non-zero at a 3 $\sigma$ level in all four frequency bands. Combined observations are counted as a single observation}
\label{tab:srcs}
\end{table}

\section{Data Analysis}
\label{sect:data}

\begin{table}
\centering
\begin{tabular}{lrr}
\hline\hline
State & Start & Stop \\
(1)      & (2)  & (3)\\

\hline
Hard & 340$^{\circ}$& 140$^{\circ}$\\
Hard Intermediate & 140$^{\circ}$ & 220$^{\circ}$\\
Soft Intermediate & 220$^{\circ}$ & 300$^{\circ}$\\
Soft & 300$^{\circ}$ & 20$^{\circ}$ \\
\hline
\multicolumn{3}{c}{{\it }}\\
\end{tabular}
\caption{Approximate angle or `hue' ranges around the power colour-colour diagram for various states. Angles are derived from a semi-major axis defined at a 45$^{\circ}$ angle to the plot axes from position 1 in Figure \ref{fig:tab}b. Definitions are based on observed power spectral shape.}
\label{tab:states}
\end{table}

We use the extensive database of RXTE observations of 12 transient Low Mass Black Hole X-ray binaries and all observations of Cygnus X-1 and Aql X-1, the selected objects and number of observations for each is shown in Table \ref{tab:srcs}. Power spectra were extracted from \xte~ PCA (Proportional Counter Array) lightcurves binned up to 1/128 s with 512 s segment size such that the lowest frequency was 1/512 s. The broad-band energy range was chosen to be that closest to 2-13 keV, dependent on the mode the observation was taken in and correcting the channels chosen for changes in the gain of the instrument over time. The lowest energy bin was excluded due to known calibration issues \citep{Gleissner04}. Scientific event modes (i.e. Event and Good Xenon modes) were preferentially chosen due to their superior time resolution, but Scientific binned modes were used when these were not available. In general measurements were taken for single observations, however both in the soft states and the low flux hard states where the variability is close to the Poisson noise level,  it was necessary to combine the power spectra for multiple observations in order to reduce the  error on the rms-variance measurements. These groupings were chosen carefully primarily on the basis of similarity in position on the HID (each group fell within a 5 \% range in hardness) and time (within 5 days of each other) but we also avoided grouping observations with obvious differences in power spectral shape or rms. 

Four broad geometrically spaced and contiguous frequency bands were selected, each band spanning a factor of 8 in frequency. The bands cover the range from 0.0039-16.0 Hz and are defined as 0.0039-0.031 Hz, 0.031-0.25 Hz, 0.25-2.0 Hz and 2.0-16.0 Hz. The variance was measured in each of these bands by integrating under the power spectra and removing the Poisson noise contribution, errors were calculated in a standard manner based on \cite{vanderKlis89} \citep[see the appendix of][for further details]{Heil12}. The Poisson noise level was estimated through fitting the white noise within the power spectra in order to reduce systematic errors caused by deadtime effects in the detector.

Comparison of the rms-Intensity diagrams in figure 4 of \cite{Heil12} reveals that sources in different states can have similar amounts of fractional rms over a broad frequency band. For this reason we choose to compare ratios of power, therefore evaluating the shape of the power spectra but removing the explicit dependence on a value describing the fractional rms which may vary from object to object. For example, the use of ratios of integrated power ensures that the effect of any constant component within the lightcurve, which will affect the observed variability amplitudes, is removed.  

We then took the ratios between the variance values in different frequency bands effectively measuring power-colours. The errors for each point were propagated in quadrature. The frequency bands are chosen to cover equal areas in log-space. Consequently, when the power spectra can be described by a power law index of 1 across the chosen frequency bands, the power-colour ratio will also be 1. The ratios were only calculated for points where the fractional variance was constrained at a 3 $\sigma$ level in all 4 frequency bands involved (these observations are classified as ``good'' in Table 1). 

\begin{figure*}
\centering
\begin{tabular}{ll}
a) & b) \\
\includegraphics[width=9.3 cm, angle=0]{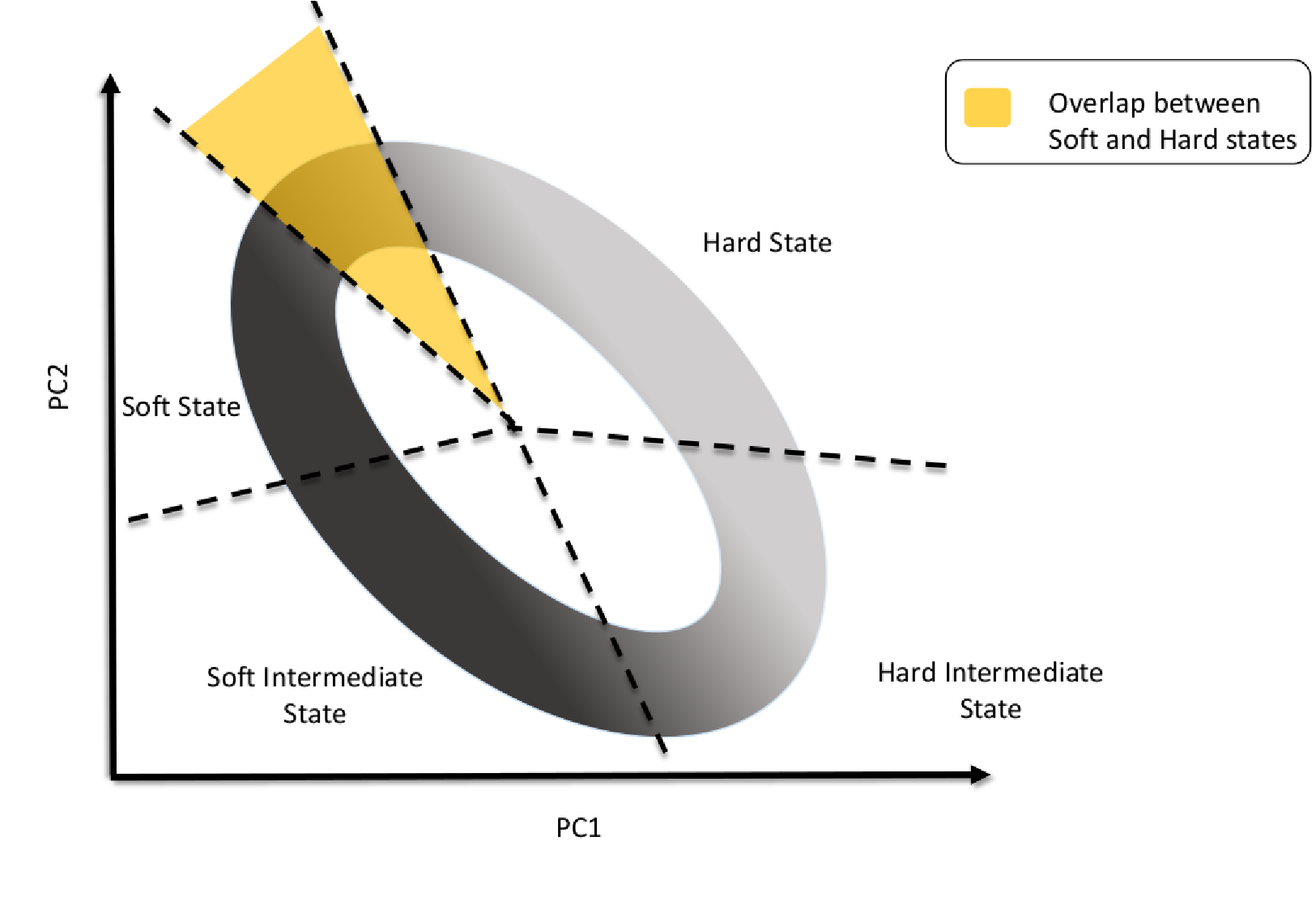} & \includegraphics[width=6.0 cm, angle=90]{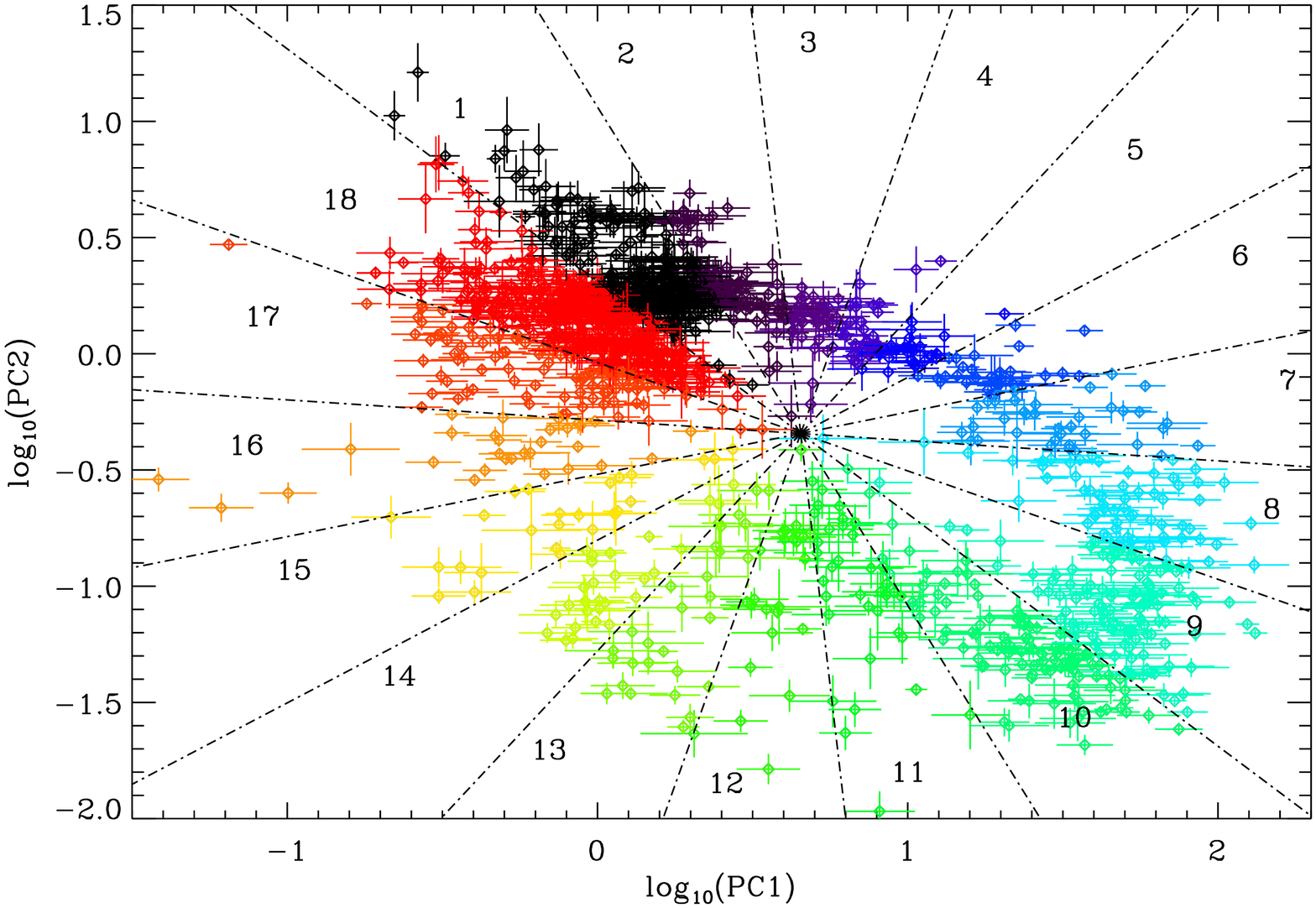}\\
\multicolumn{2}{l}{c)}\\
\multicolumn{2}{c}{\includegraphics[width=12.0 cm, angle=90]{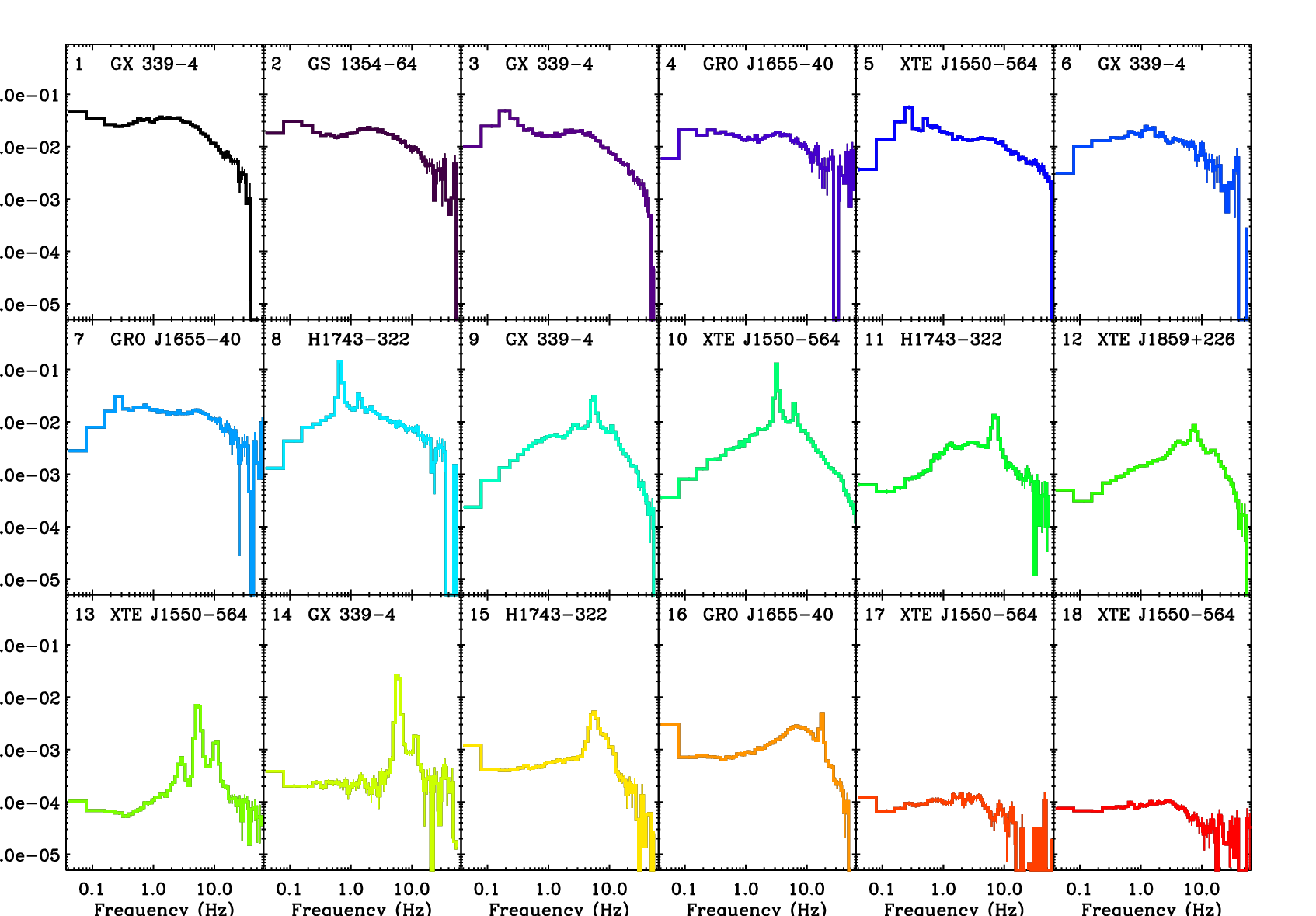}}\\
\end{tabular}
\caption{\emph{a:} Graphic illustrating where various states appear within the power-colour diagram. The area of overlap between the hardest and softest states is also indicated. \emph{b:} Power colour-colour plot for all observations of the transient objects within the sample with labels indicating 20-degree azimuthal or `hue' regions from which the power spectra given in \emph{c} were found. The plot is colour-coded for each 20$^{\circ}$ bin with the same colours used in \emph{c}. \emph{c:} Example power spectra for each of the 20 degree ranges of hue around the power colour-colour diagram. Colours and indices refer to the 20$^{\circ}$ angular bins used in \emph{b}. Further examples are given in the Appendix.}
\label{fig:tab}
\end{figure*}

\section{Results}
\subsection{Power-colour Analysis}

The power colour-colour plot is a quick, simple way to evaluate the shape of the power spectra without requiring detailed power-spectral fitting. We go on to demonstrate that it has the unique property of showing a well constrained shape for all BH LMXBs allowing easy state classification for new sources with only a limited number of observations required.

Figure \ref{fig:srccols} shows the power colour values for two particular frequency ratios, Power colour ratio 1 (PC1) is defined as variance in 0.25-2.0 Hz / 0.0039-0.031 Hz and ratio 2 (PC2) is variance in 0.031-0.25 Hz / 2.0-16.0 Hz. These particular ratios not only compare all four broad frequency bands used in the initial analysis, making the most of the available data, but are also separated in frequency. This plot is colour coded according to object, and the similarity in power-spectral evolution throughout outbursts between different sources can be clearly observed through the similarity in tracks traced. Log axes are used in order to highlight small changes in power spectral shape during source evolution.  

The evolution of objects around the power colour-colour diagram during an outburst takes the form of a loop and can be simply described in terms of a single dimension by measuring an azimuthal coordinate of each data point (plotted in terms of the logarithm of the power-colours), relative to some central point.  The central point used is the mid point of the x and y extent of the plot, with position [4.51920, 0.453724] in linear units.  The semi-major axis is defined at a 45 degree angle to the x and y axes.  By analogy with the  `colour-wheel' representation of colours, we call this angle around the power colour-colour plot the ``hue''.  To simplify the classification we group the power-spectra according to their hue, splitting the hue angles into 18 sectors, each corresponding to 20 degrees of arc which are indexed on Figure \ref{fig:tab}b, example power spectra for each segment are given both in Figure \ref{fig:tab}c and the Appendix. The ranges of hue position for the different states are given in Table \ref{tab:states}. For an in depth discussion of power spectral evolution for multiple objects over the course of an outburst see \cite{KleinWolt08}.

\begin{figure}
\begin{center}

	\includegraphics[width=6.0 cm, angle=90]{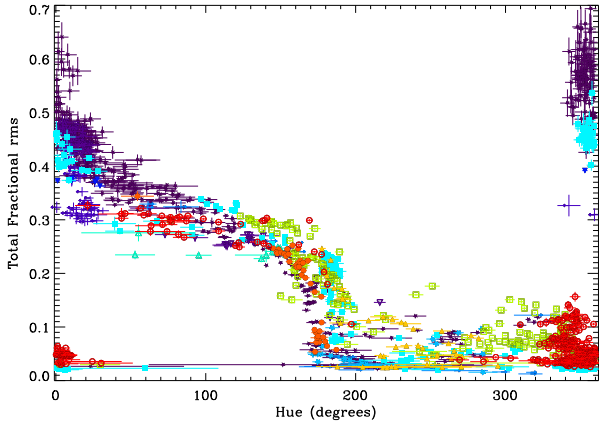}
\end{center}
\caption{Total fractional rms within the 0.0039-16.0 Hz range vs. power-spectral hue for all observations in the sample. Colours and symbols are identical to those used in Figure \ref{fig:srccols}. Some separation in hue can be observed between observations with identical fractional rms, this separation is linked to the strength of the low-frequency quasi-periodic oscillations observed from these objects which differs widely.}
\label{fig:rmshue}
\end{figure}

Defining the radial central point in the power-colour diagram in an objective way is difficult due to the radial scatter and relative sampling of different states, with most of the measurements being obtained from the longer lasting `pure' hard and soft states which populate the top left corner of the diagram.  Thus the main criterion we use for defining this radial point is practicality.  Firstly, the position of the radial centre should be relatively free of data points, so that a continuous progression in timing properties follows a continuous progression in hue rather than jumping sharply between disconnected values of hue.  Secondly, the radial centre should not be so far from the hard and soft-state data points that they blur completely together into just a narrow range of hue.  We find that our definition of the radial centre satisfies these constraints well.  Slightly different, but still reasonable, values of the radial centre (up to 0.2 in log(power-colour) in the horizontal and 0.1 in the vertical) give slightly different values of the hue for the observations, but also lead to the same key results:  namely, that the the drop in rms observed during the transition through the hard intermediate states occurs at approximately the same hue for all objects and that the transition in energy-spectral hardness from the hard to the soft state is well defined for all objects and also occurs at the same hue (see Figure \ref{fig:rmshue} and Heil, Uttley \& Klein-Wolt 2015 for further discussions of these points).  We also note that the average error on the hue for the observations is 5$^{\circ}$, larger than the differences expected from small changes in the assigned central point.

For some observations there may be variation in power-colour within the observation \citep[see i.e][]{Homan01, Motta12, Heil12}. This is particularly likely to affect the intermediate states, with strong frequency shifting or particular components appearing/disappearing rapidly within an observation. In this case there may be some intrinsic variation which is not fully captured by the current error bars on points within the power-colour-colour plot. However, by testing for intrinsic variation in shape within a large sample of single observations, \cite{Heil12} demonstrated that this only appears to affect a small sub-set of power spectra from BHBs. Whilst we could have measured single power colour points per orbit rather than per observation in order to limit this effect, this lowers the signal to noise and thus increases the errors on each point. These points are then typically coincident in the diagram anyway, so we choose to continue averaging the power spectra over all orbits within an observation.

We now go on to discuss the position of observations in specific states within the power colour-colour diagram with particular reference to how the gradual evolution of the power spectra causes a loop around the parameter space.

\emph{Hard States:}  In this state the variability is strong and visible over a wide range of frequencies, as such the power spectra consists of very broad approximately flat-topped noise, an example of which is shown in Figure \ref{fig:tab}c plot 1. Due to the broad nature of the noise, the power-colour ratios are both approximately 1.0. Consequently, the points appear in the upper left hand corner of the power-colour diagram. This is illustrated by the yellow shaded region on Figure \ref{fig:tab}a. As the flux from the source increases as the outburst progresses, the strongest variability becomes concentrated into a smaller frequency range, this behaviour can be seen in power spectra 2-7 in Figure \ref{fig:tab}c. Due to the more peaked nature of the broad-band noise objects transition along the light grey path illustrated in Figure \ref{fig:tab}a until they reach the HIMS, we note that there is gradual evolution in the power spectra from the hard state into the HIMS with no dramatic differences between the two. The difference is more clearly observed in the energy spectra, which now start to become softer.

\emph{Hard Intermediate States:} In the HIMS (PSDs 8-11 in Figure \ref{fig:tab}c) the power spectra reach their most strongly peaked shape, with most power concentrated in the two highest frequency bands. Due to the fact that this is the most strongly concentrated that the broad band noise becomes, these observations can be found in the bottom right-hand corner of the power colour-colour diagram. The type C quasi-periodic oscillations are particularly strong within this state and, as they are found with a range of different amplitudes in various objects, they can likely explain some of the spread which can be observed in the power colour-colour diagram in this state. We explore this point further in \cite{Heil14}.

\emph{Soft Intermediate States:} The transition to the SIMS is characteristically only strongly observed in the variability from the source. The peaked broad-band noise disappears, leaving only type-B QPOs and weak low-frequency power-law-like noise. There is huge variation in variability properties within this state, some of which is illustrated in the power spectra for positions 12-15 in both \ref{fig:tab}c and the Appendix. Evolution within the SIMS is not straightforward. Sources may pass through the SIMS multiple times, transitioning into a soft state or returning to a hard-intermediate state. Consequently objects do not evolve around the SIMS part of the loop marked out in Figure \ref{fig:tab}a in a linear fashion, but typically pass through this region very rapidly as they transition. The variety in power spectral shapes within the SIMS explains the wide distribution of points in the power colour-colour diagram within this region. However, we can still use the power colour-colour diagram to interpret what individual observations may look like based on their measured hue. All observations in this state contain a strongly peaked high frequency component with the majority of the power above $\sim$ 1 Hz, they also contain a weak power-law like component at low frequencies. The stronger this low-frequency component the further clockwise observations can be found on the power colour-colour diagram. Figure \ref{fig:tab}c illustrates this point, moving from position 13 to 15 a large difference in relative normalisations between the low and high frequency components can be observed.

\emph{Soft states:} The soft states typically show low frequency variability with an approximate power-law of -1 bending to a steeper power-law at high frequencies ($\sim$ 5 Hz), examples can be found in plots 17 and 18 in Figure \ref{fig:tab}c. Although the strength of the variability in this state is much lower than that of the hard states ($\sim$40$\%$ vs. $<$3$\%$) the power-colour method removes any dependence on power spectral normalisation, consequently most observations in the soft states appear in an identical area of the power colour-colour diagram to the hardest hard states (Figure \ref{fig:tab}c plot 1). This overlap is in the yellow shaded region on Figure \ref{fig:tab}a. The states of these observations can be easily separated by either total rms or hardness and consequently their true states are easily identifiable. 

\begin{figure}
\begin{center}

	\includegraphics[width=6.0 cm, angle=90]{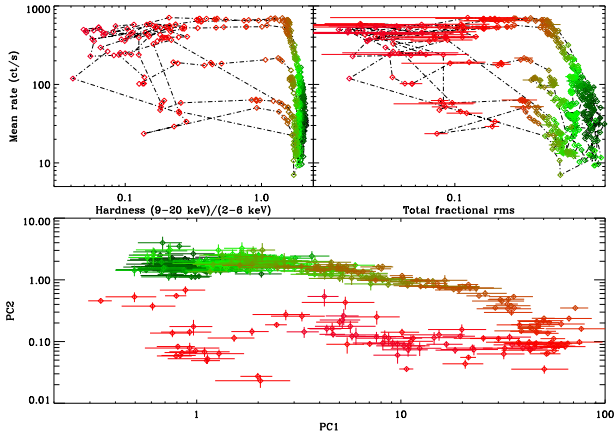}
\end{center}
\caption{HID, RID and Power colour diagram for GX 339-4 for all `good' observations included within the sample. The total fractional rms is measured in the 0.004-16.0 Hz band as used for the power colour-colour diagram the full range of states is clearly visible as it follows a well defined track around the power colour-colour diagram.}
\label{fig:GX339path}
\end{figure}

Once objects begin the return back to the hard state they follow an identical path around the power colour-colour diagram but in an anti-clockwise direction. This is demonstrated in Figure \ref{fig:GX339path} where all observations follow a very well constrained path both into and out of outburst. Simply put, objects initially transition around the power colour-colour diagram in a clockwise direction from position 1-18 in Figure \ref{fig:tab}b, and then return back towards quiescence in an anti-clockwise direction (although the exact path found for each object is clearly dependent upon the range of states sampled by the observations during the outburst). 

\begin{figure}
\begin{center}

	\includegraphics[width=6.0 cm, angle=90]{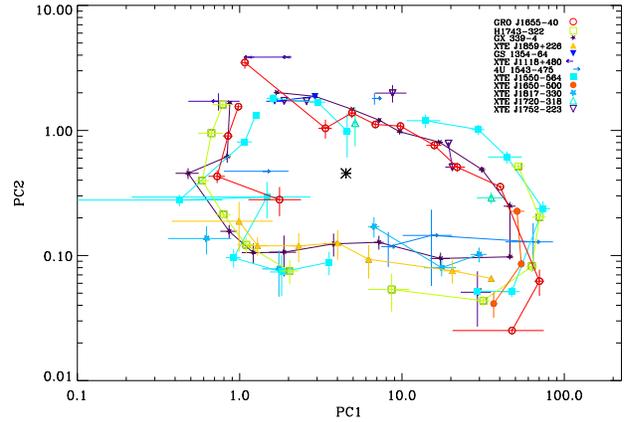}
\end{center}
\caption{Power colour-colour plot binned according to the hue position around the curve for each object.  Points in adjacent angular bins are connected and the central point for the hue calculation is marked.}
\label{fig:angbin}
\end{figure}

Other frequency bands or ratios will clearly trace different paths around the power-colour-colour plot. Not all paths are necessarily loops, but these particular bands have been selected in order to sample the broad-band changes in the power spectra of black hole X-ray binaries using observations taken with RXTE. The advantage in the ratios used here is that almost all states can be cleanly separated, which is not necessarily the case when other frequency bands are used. In addition we note that the power spectra observed from black hole X-ray binaries are known to show some dependence on energy, this is particularly visible in the intermediate states, with high frequency variability being stronger at harder energies and some low frequency noise stronger in the soft bands \citep[see i.e. figure 6][]{Gierlinski05, Wilkinson09}. As such the power-colour-colour plot will show some dependence on the energy band taken. The current energy band used (2.5-13 keV) is fairly broad, but narrower soft bands may show some shift, for instance in a harder energy band power spectra in the HIMS become even more strongly peaked and consequently the points around 180 degrees of hue will shift both farther to the right and down. 

Using the assigned 20 degree hue ranges we can also bin the power colour-colour diagram according to source in order to test how similar the power colour-colour paths are for different objects. It also allows us to identify whether the spread observed in the lower-left-hand corner is purely due to a spread between different objects. Figure \ref{fig:angbin} shows the results for all transient black holes studied. Although single objects show significant `radial' scatter in the power-colour-colour diagram their mean paths are clearly similar with most points falling within 3 sigma of each other (for the standard error on the mean). There are some notable differences, XTE J1550-564 and H1743-322 appear to follow a different path out of the hard state than other objects. In \cite{Heil14} we demonstrate that this deviation is caused by the particularly strong QPOs present in the lightcurves from these sources, direct comparison of the power spectra at positions 6,7 and 8 (see both Figure \ref{fig:tab}c and the Appendix) shows that these features are clearly much stronger in these objects in this particular state than in GX 339-4. \cite{Motta14} demonstrate that this difference appears to be due to the inclination angles of the sources, XTE J1550-564 and H1743-322 have higher inclinations than GX 339-4 and stronger QPO amplitudes.

We can also see this deviation, caused by the QPOs, in Figure \ref{fig:rmshue}. In this plot we demonstrate the true power of the power colour-colour diagram in that it allows us to directly compare the parameters of observations with similar power spectral shape in a model independent manner. Figure \ref{fig:rmshue} shows that there is a slight shift in hue for some observations of XTE J1550-564 and H1743-322 between $\sim$150 and 200 degrees, although this shift is not universally observed for all observations. We explore this point in more detail in \cite{Heil14}.

Although outbursts of individual sources follow constrained paths around the power colour-colour diagram, overall the paths are very similar and the power colour-colour diagram is strikingly consistent for a wide range of objects. It is already well known that the evolution of characteristic frequencies in the power spectrum is highly consistent both between different outbursts of the same source and outbursts of many objects \citep{KleinWolt08}.  However, the similarity in power-colour evolution further demonstrates {\it in a model-independent way} that the relative frequencies {\it and} amplitudes of power-spectral components must vary in a consistent way.

\subsection{Outlying points}

 \begin{figure}
\begin{center}

	\includegraphics[width=6.0 cm, angle=90]{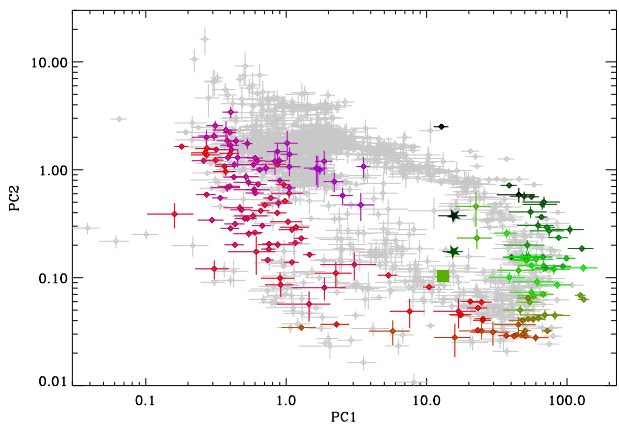}
\end{center}
\caption{Power colour diagram for H1743-322 for all `good' observations included within the sample with observations known to show unusual properties highlighted. Observations are colour-coded according to hardness going from green-red as the source evolves from hard to soft. Stars show points with additional low frequency QPOs and the square is an observation with occultation dips indicating that this is a high inclination object.}
\label{fig:1743odd}
\end{figure}

Since the paths followed by all objects as they transition around the power colour-colour diagram are so consistent we should be able to identify unusual observations as they are likely to be ``out of place'' relative to i.e. other points of similar spectral hardness. In order to test this hypothesis, we identify the position of three observations of H1743-322 within its power colour-colour diagram, two of these observations have very unusual low frequency ($\sim$ 0.01 Hz) QPOs \citep{Altamirano12}, which have only been observed from this source, and the third is known to show a hard dip \citep{Homan05}, which is associated with the high inclination of the system to our line of sight. In all of these cases there is additional power in the lowest frequency band relative to other observations in a similar state. Figure \ref{fig:1743odd} shows the power colour-colour diagram of H1743-322 with these points identified. As this object is relatively faint there are few observations when the power in well constrained in all bands during the low-hard state, so the full evolution is not visible. However the colour-coding shows that these three observations are clearly separated from their counterparts with similar measures of hardness, the value of the ratio PC1 (containing the lowest frequency band) is smaller. This demonstrates that the power colour-colour diagram can be used to identify power spectra with unusual features without the detailed fitting and investigation often required to find them when they would not be easily identifiable using other methods \citep[see][for a discussion of the points with low frequency QPOs within the HID]{Altamirano12}. 

\subsection{Exploring the comparison of other object types to the black hole transients}

Power colours can be measured for a range of different accreting objects, from neutron star X-ray binaries and cataclysmic variables to Active Galactic Nuclei. The consistency observed in the power colour-colour diagram for the transient low mass black holes suggests that it may be a suitable method to classify new objects, particularly given the low levels of signal to noise required to measure a power-colour ratio. We now go on to make initial explorations into how other X-ray binary types' power colour-colour diagrams compare to that of the transients. Specifically we make comparisons to a high mass black hole X-ray binary Cygnus X-1 and an atoll type low mass neutron star X-ray binary Aquila X-1.

\subsubsection{Comparing the transient black holes to Cygnus X-1}

\begin{figure}
\begin{center}
	\includegraphics[width=6.0 cm, angle=90]{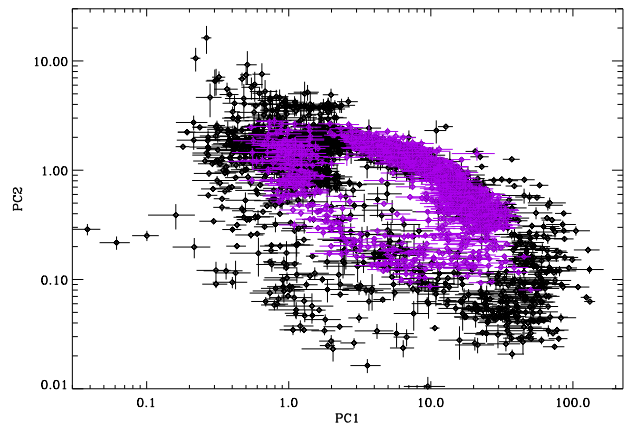}
\end{center}
\caption{All observations of Cygnus X-1 (purple) and the transient black holes (black) the similarity in power colour-colour evolution between the two object types is clearly visible.}
\label{fig:CygX1powcol}
\end{figure}

Figure \ref{fig:CygX1powcol} shows the power colour-colour plot for all observations in the LMXB sample (black), with the power colours for the black hole high mass X-ray binary (HMXB) Cygnus X-1 over-plotted (purple). The example power spectra from observations of Cyg X-1 are shown in the Appendix. The x-axis extent of the power colour-colour path for Cyg X-1 is slightly different to that of the transients (discussed further below), however the overall shape is very similar.

The power colour-colour plot of Cyg X-1 appears to indicate that the power spectra of this source show all states observed in the LMXBs except the most extreme low-flux hard states with very broad noise. As such it is possible to discern between the hardest and softest states at the top of the power colour-colour diagram. Energy spectral analysis suggests that Cyg X-1 does not appear to ever reach a true soft state \citep{Belloni10}, where the energy spectrum is dominated by a disc blackbody component and only a weak power-law tail is present. However the power colour-colour diagram demonstrates that the shapes of the power spectra are similar to those observed from the LMXBs (see power spectra from regions 17 and 18) even though the total variance (i.e. the normalisation of the power spectra) is higher for Cyg X-1 in the softest states. It is clear that the PC2 ((0.031-0.25 Hz)/(2-16 Hz)) power-colour of Cyg X-1 rarely reaches values greater than 1.5. This is in contrast to the transient objects, which show observations with ratio values up to 2.3 in the hard state and 3.7 in the soft. Upon further examination we find that the soft state observations from the LMXBs with y-extent values on the power colour-colour diagram greater than those observed in the soft-state observations of Cyg X-1, are all from an unusual state of GRO J1655-40 which is not observed from Cyg X-1, they are discussed further in Uttley \& Klein-Wolt (in prep.). Since Cyg X-1 does not reach the broad low luminosity hard states observed from the BH transients  we neglect these states from the discussion below. This point also illustrates why the power colour-colour loop for Cyg X-1 begins at a larger hue than the transient objects, there are very few in the hard state with the very broad-band noise observed in PSDs from position 1 in Figure \ref{fig:tab}b.

Although the shape of the power colour-colour loop followed by Cyg X-1 is similar to that of the transient black holes we observe that there is some deviation between the parameter space covered by the two loops in the HIMS. The LMXB points are clearly delineated from those of Cyg X-1, moving further down and to the right. In order to create this deviation either the variance at low frequencies must be higher in Cyg X-1 than the transients or there must be higher variance in the LMXBs at high frequencies. As these are the states in which the QPOs are strongest and are in fact the dominant feature within the power spectra, the obvious explanation is that this deviation is primarily caused by the extra variance present in these coherent features, which are not visible in the lightcurves from Cyg X-1. However inspection of the power spectra reveals that this is not necessarily the complete explanation. Comparing the power spectra of Cyg X-1 in locations 10, 11 and 12 to those of the transients reveals that when the amplitude of the broad band noise under the QPOs in the highest two frequency bands is similar to that of Cyg X-1, there is consistently more power at low frequencies in Cyg X-1 than in the transients. \cite{Axelsson05} found that an additional power law was required at low frequencies in order to fit the power spectra of Cyg X-1 in these states, indicating that the deviation between the power colour-colour path of Cyg X-1 and the transients is likely to be due to the stronger low-frequency power-law noise in Cyg X-1 together with the presence of the QPOs in the LMXBs.  

\subsubsection{Comparing the transient black holes to Aql X-1}

\begin{figure}
\begin{center}
	\includegraphics[width=8.0 cm, angle=0]{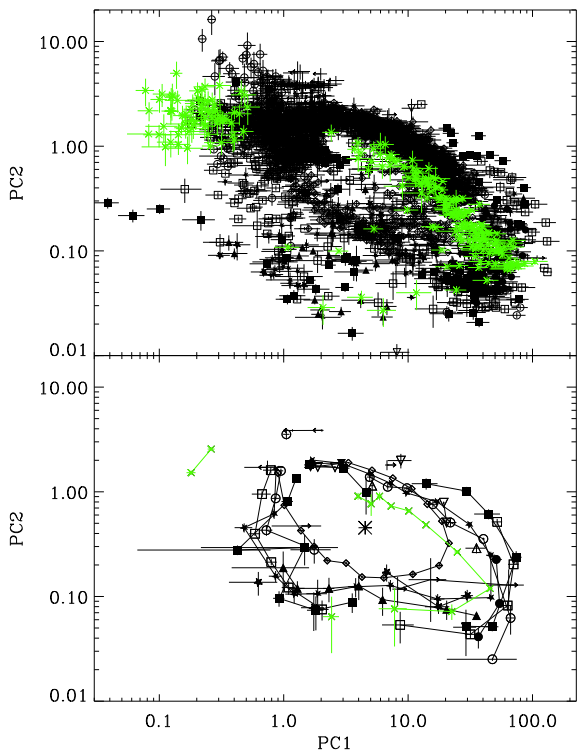}
\end{center}
\caption{All observations of the black holes (black) (including Cygnus X-1) and Aquila X-1 (green). The lower plot shows the same points binned at 20 degrees of hue. Symbols are the same as those used in Figure 1, note that the binned Cyg X-1 points are illustrated with diamonds. The central point for the hue calculation is marked on the binned plot.}
\label{fig:Aqlcomp}
\end{figure}

It has been shown in previous work that the power spectra measured from observations of neutron stars show distinct similarities to those of the black holes \citep{Wijnands99, Sunyaev00, Belloni02, vanderKlis06, KleinWolt08}, with the neutron stars showing an additional high frequency component extending the broad-band noise above 100 Hz. We therefore take all observations of the atoll source Aql X-1 to make a brief comparison between this object and the black hole X-ray binaries. Figure \ref{fig:Aqlcomp} shows that the states of the neutron star cover a similar region in the power colour-colour diagram to the transient black holes (and indeed Cyg X-1), but part of the path is shifted. The observations of Aql X-1 in the island (IS), lower-left banana (LLB) and lower banana (LB) states correspond to the hard path followed by the black holes, only translated downwards on the power colour-colour plot, the translation is particularly clear on the binned version of the plot (lower panel). This is due to extra 2--16~Hz noise present in the neutron stars, a difference first identified in power-spectral comparisons by \citet{Sunyaev00}. At first glance the softer states (upper banana; UB) from the atoll source, correspond closely on the power colour-colour diagram with those observed from the black holes, however comparison between objects reveals that most of the overlapping observations are either from anomalous soft-state observations of GRO J1655-40 (discussed in Uttley \& Klein-Wolt in prep). Although they are dominated by power-law like noise at low frequencies (\textless 0.1 Hz) in a similar manner to the soft state black hole observations, they show a drop in power above $\sim$1 Hz. Consequently some observations in this state do not appear within the sample, as there is no power measured above the noise in the highest frequency band. 

The differences between the black hole X-ray binaries and the neutron star Aql X-1 in the power colour-colour diagram suggest that this diagram can be used as a diagnostic tool, aiding in identifying whether newly identified objects are neutron stars or black holes in a fast manner.

\section{Conclusions}
We have demonstrated a simple, model-independent method for evaluating the shape of power spectra. By comparing power-colours -- ratios of integrated power (rms-variance) in different frequency bands -- we can evaluate how the power spectra evolves between different states. This method also allows for direct comparison between different objects independent of the total amplitude of the rms-variability within the lightcurve. We have found that:
\begin{itemize}
\item{The power colour-colour diagram shows that, across the LMXBs the power spectra evolve over an outburst in a very similar manner.  If the power-spectra are represented in terms of multiple components (e.g. \citealt{KleinWolt08}), these must evolve similarly in both frequency {\it and} amplitude.}
\item{The general evolution can be simply parameterised in terms of the angle along the power-colour-colour track, which we call the `hue'.  For a given power-spectral hue there remains significant scatter in power-colours in the intermediate states for a given object, but the mean power-colours are similar for different objects, again indicating a common evolution.}
\item{We can use the power colour-colour diagram to identify unusual or `out of place' observations.  These include high inclination objects with dips in their power spectra and additional low frequency features.}
\item{The overall shape of the power-colour-colour track shown by the persistent black hole HMXB Cyg~X-1 is similar to that seen in the transient LMXBs.  Cygnus X-1 shows nearly all the broad power-spectral shapes observed in the LMXBs excepting those seen in the low luminosity hard states which Cyg X-1 never reaches.}
\item{The power-colours of Cygnus X-1 differ from those observed in the LMXBs when the objects are in hard intermediate states.  The difference is due to a combination of stronger low-frequency power-law noise in the Cyg~X-1 power spectra, combined with the presence of strong QPOs in the LMXB power spectra.}
\item{We find that parts of the power colour-colour diagram of the Atoll source Aql X-1 differs from that of the black holes, suggesting that this method could provide a simple new tool to help identify the nature of newly discovered X-ray binaries.}
\end{itemize}
The power-colour and power-spectral hue is a purely empirical classification, but due to its simplicity and model-independence, it should prove extremely useful in comparing timing evolution with other observables, which will help to unlock the physical origin of the different states.  Furthermore, the broad frequency bands used to determine the power-colours can be used for fainter sources where detailed power-spectral measurements may not be possible.  Assuming that the Poisson noise power dominates the source power in this range, the minimum rms-squared variance for a 3$\sigma$ detection of variability in a frequency band of width $\Delta \nu$ in an observation of length $T_{\rm obs}$ is:
\[
\sigma^{2}_{\rm min}=\frac{6(S+B)}{S^{2}}\sqrt{\frac{\Delta \nu}{T_{\rm obs}}}
\]
where $S$ and $B$ are the source and background photon count rates respectively.  Thus for the 2--16~Hz band, where we expect the lowest signal-to-noise, we expect in 10--100~ks to detect variability from sources through the hard and intermediate states, down to count rates $\sim10$~count~s$^{-1}$ (assuming negligible background).  This will enable the power-colours used here to be measured for bright X-ray binaries in nearby galaxies with current imaging X-ray telescopes.  It would also be interesting to extend this technique to study other types of variable object.

\section*{Acknowledgements}
We would like to thank Jeremy Heyl for pointing out the analogy between the azimuthal angle in the power colour-colour diagram and the hue used in RGB colour classification. We also thank the referee for their helpful comments and suggestions.


\bibliographystyle{mn2e}
\bibliography{powcolini.bib}

\bsp
\appendix
\section{ }
Example power spectra for each of the 18 positions illustrated in Figure \ref{fig:tab}b. Power spectra were chosen randomly from the observations of transient black holes in order to illustrate the range of power spectral shapes observed from the black hole X-ray binaries. Power spectra of Cyg X-1 are also shown for each position. In the position where hard and soft states overlap (18) hard state spectra are labeled HS and soft state SS.  

\begin{figure*}
\begin{center}$
\begin{array}{ll}
	\includegraphics[width=6.0 cm, angle=90]{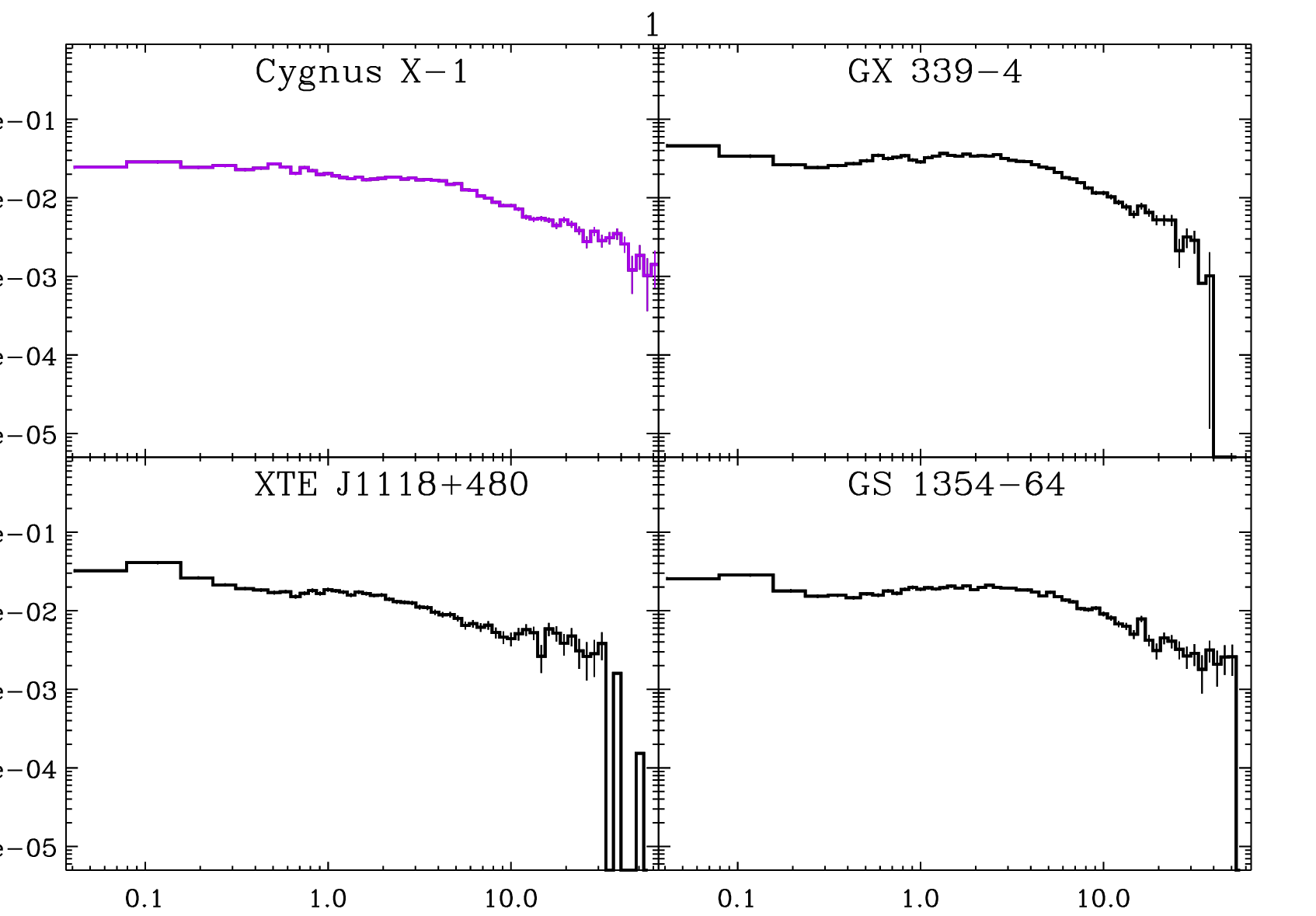} &
	\includegraphics[width=6.0 cm, angle=90]{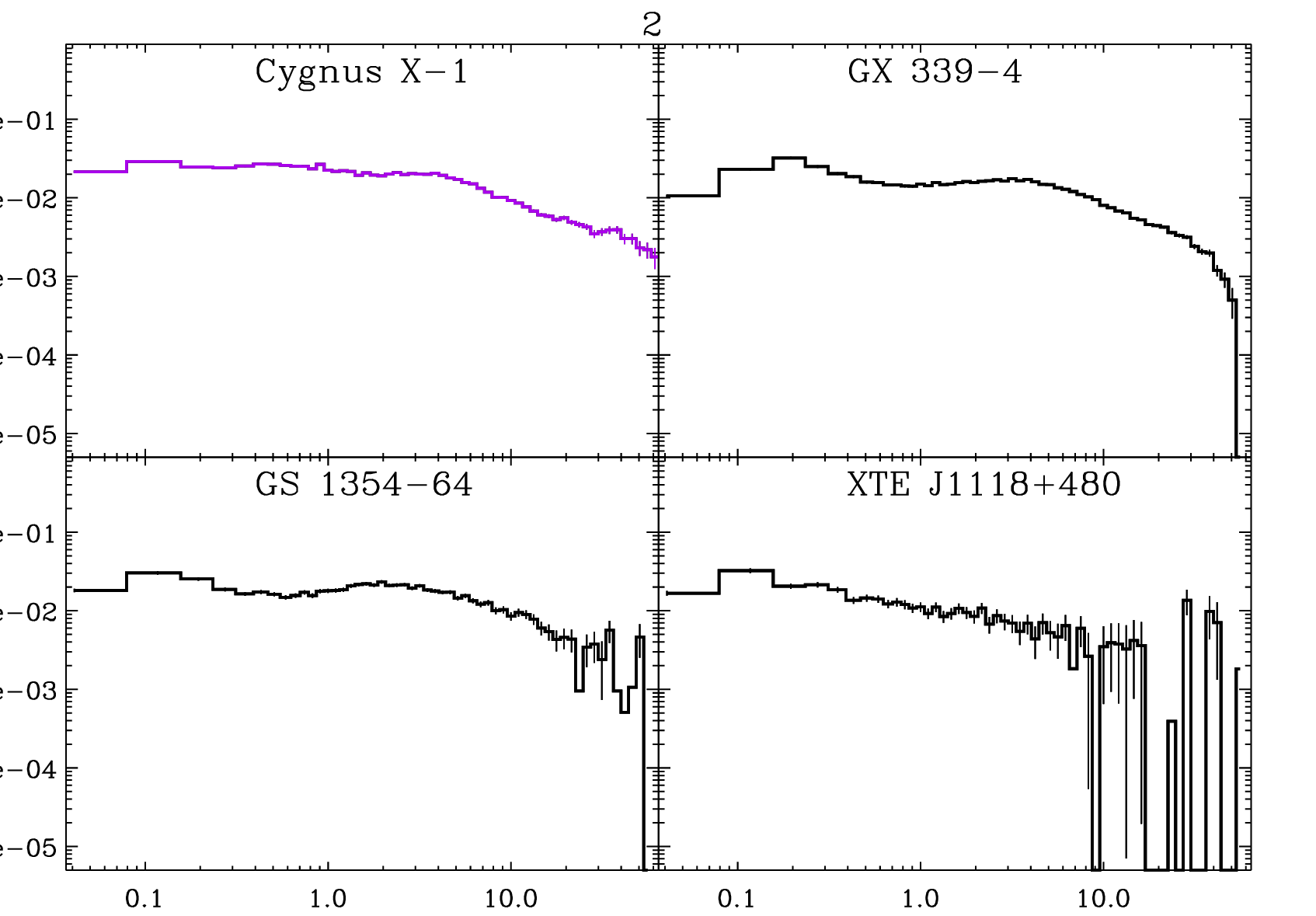}
\end{array}$
\end{center}
\end{figure*}

\begin{figure*}
\begin{center}$
\begin{array}{ll}
	\includegraphics[width=6.0 cm, angle=90]{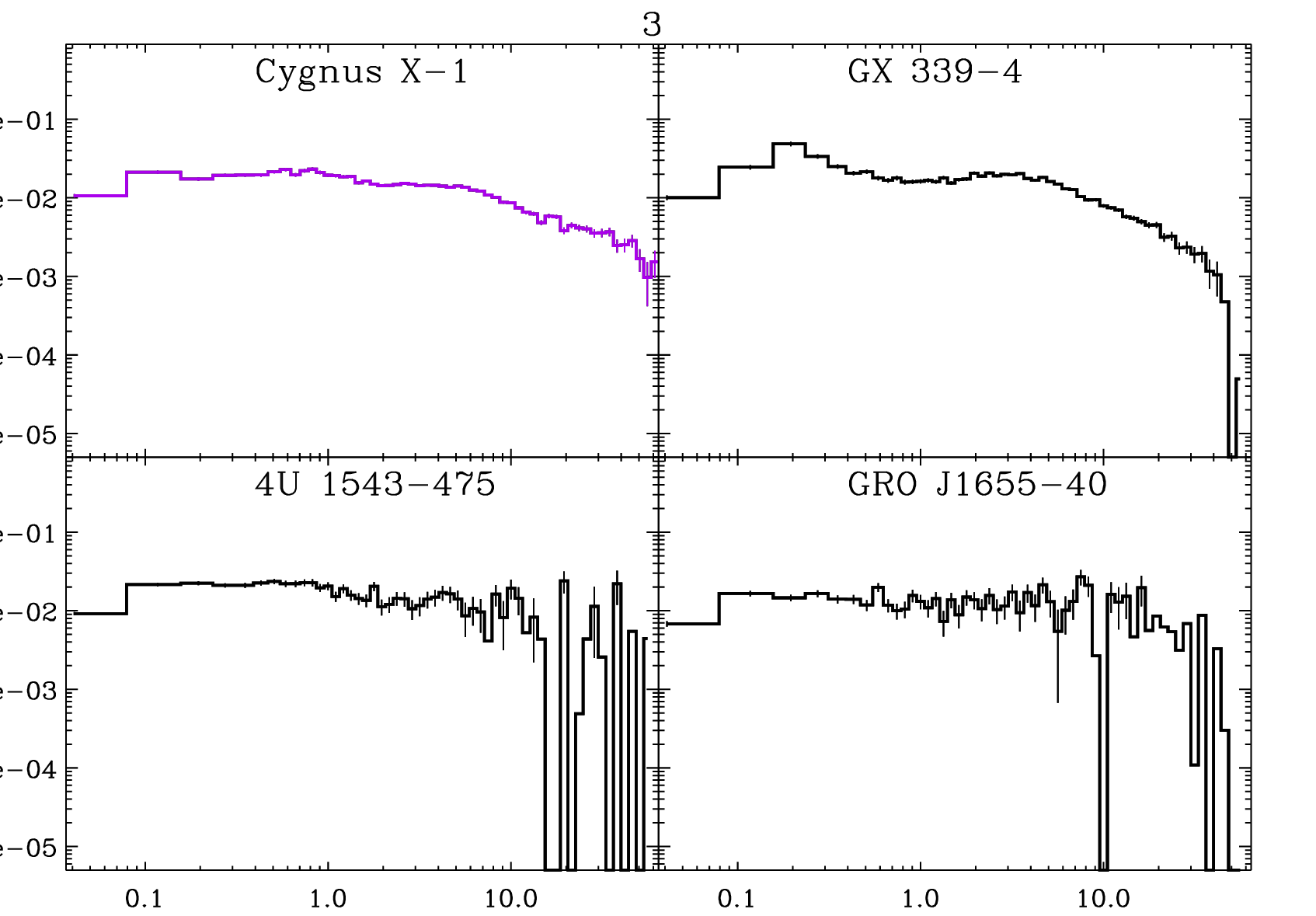} &
	\includegraphics[width=6.0 cm, angle=90]{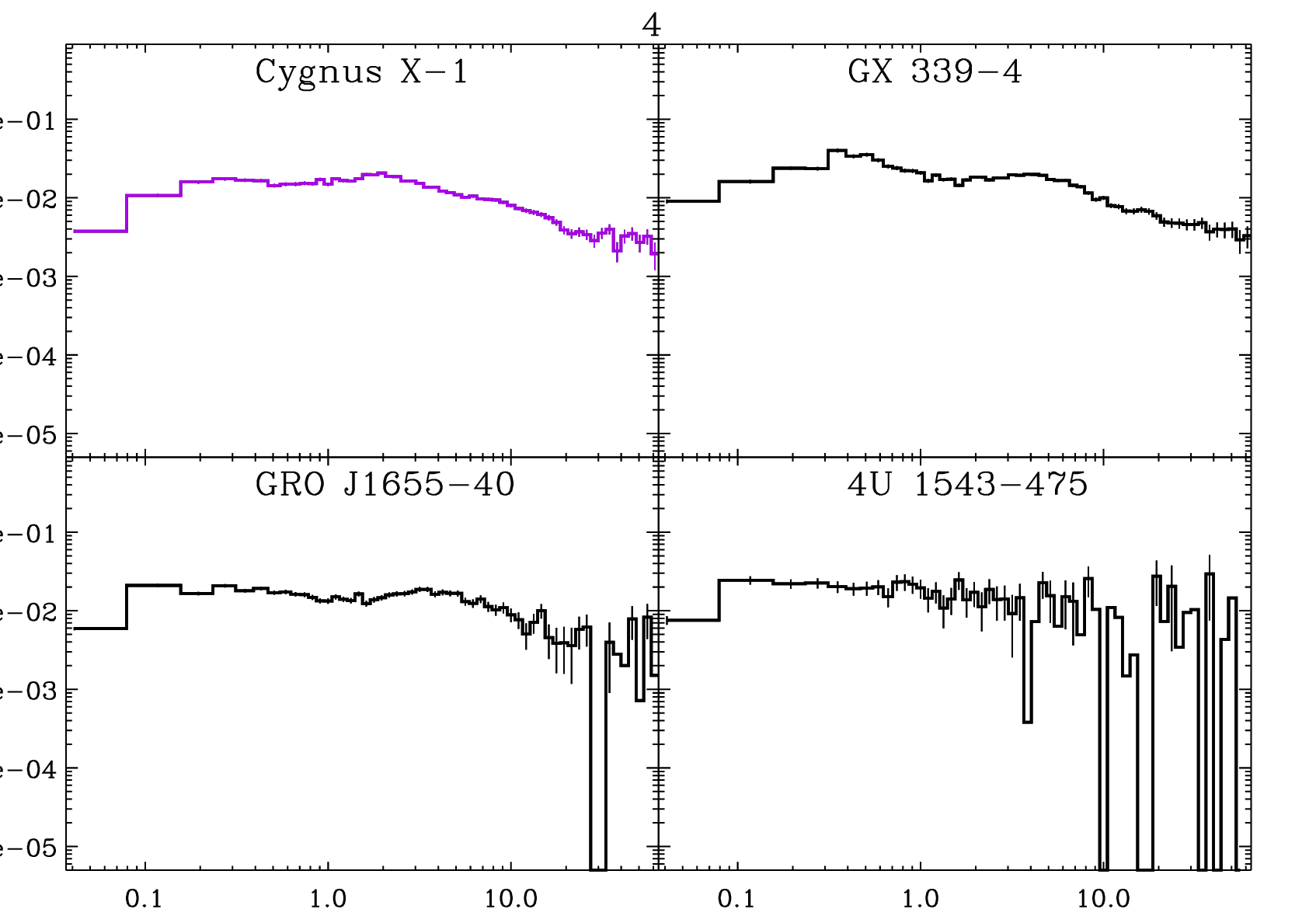}
\end{array}$
\end{center}
\end{figure*}

\begin{figure*}
\begin{center}$
\begin{array}{ll}
	\includegraphics[width=6.0 cm, angle=90]{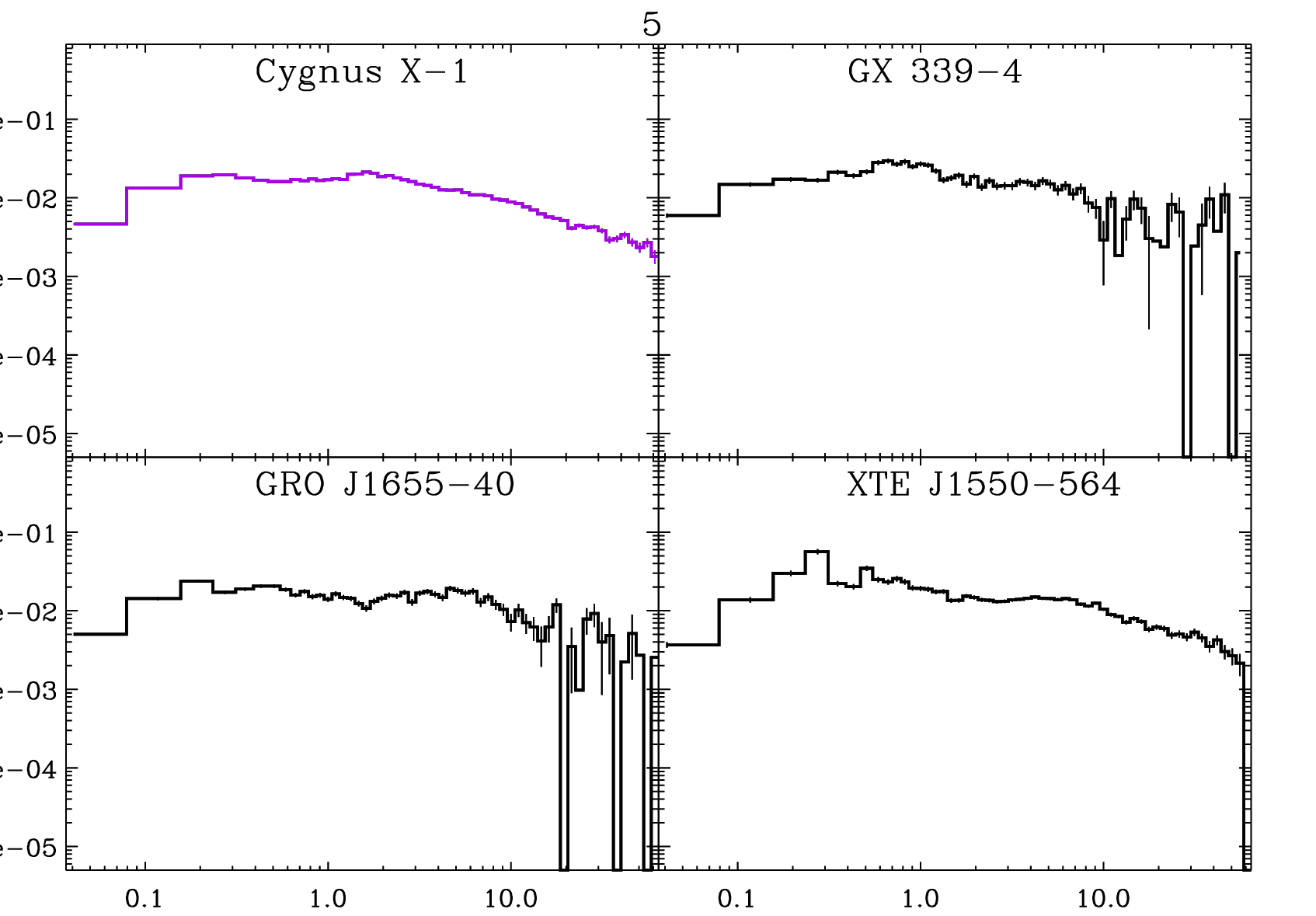} &
	\includegraphics[width=6.0 cm, angle=90]{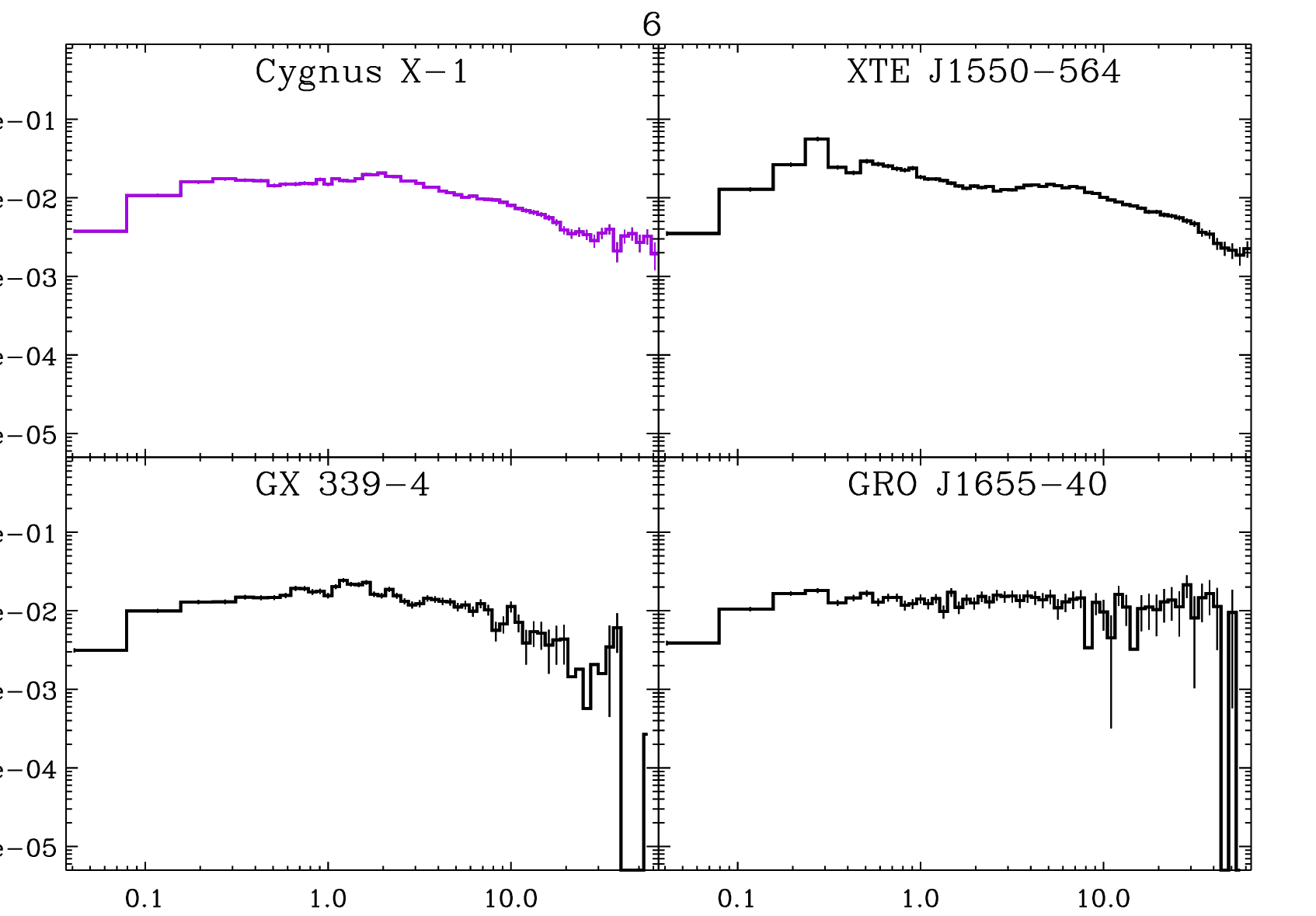}
\end{array}$
\end{center}
\end{figure*}

\begin{figure*}
\begin{center}$
\begin{array}{ll}
	\includegraphics[width=6.0 cm, angle=90]{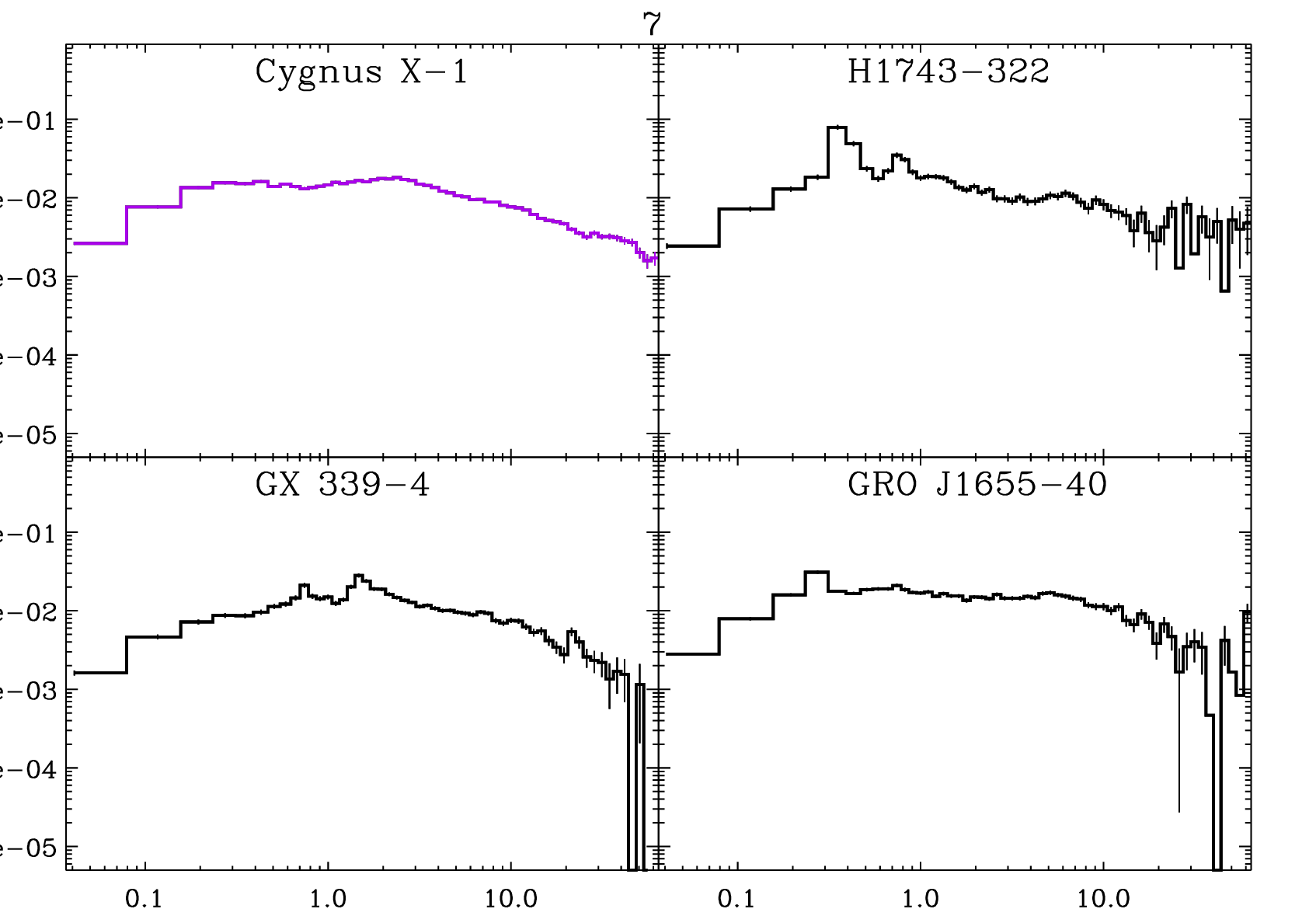} &
	\includegraphics[width=6.0 cm, angle=90]{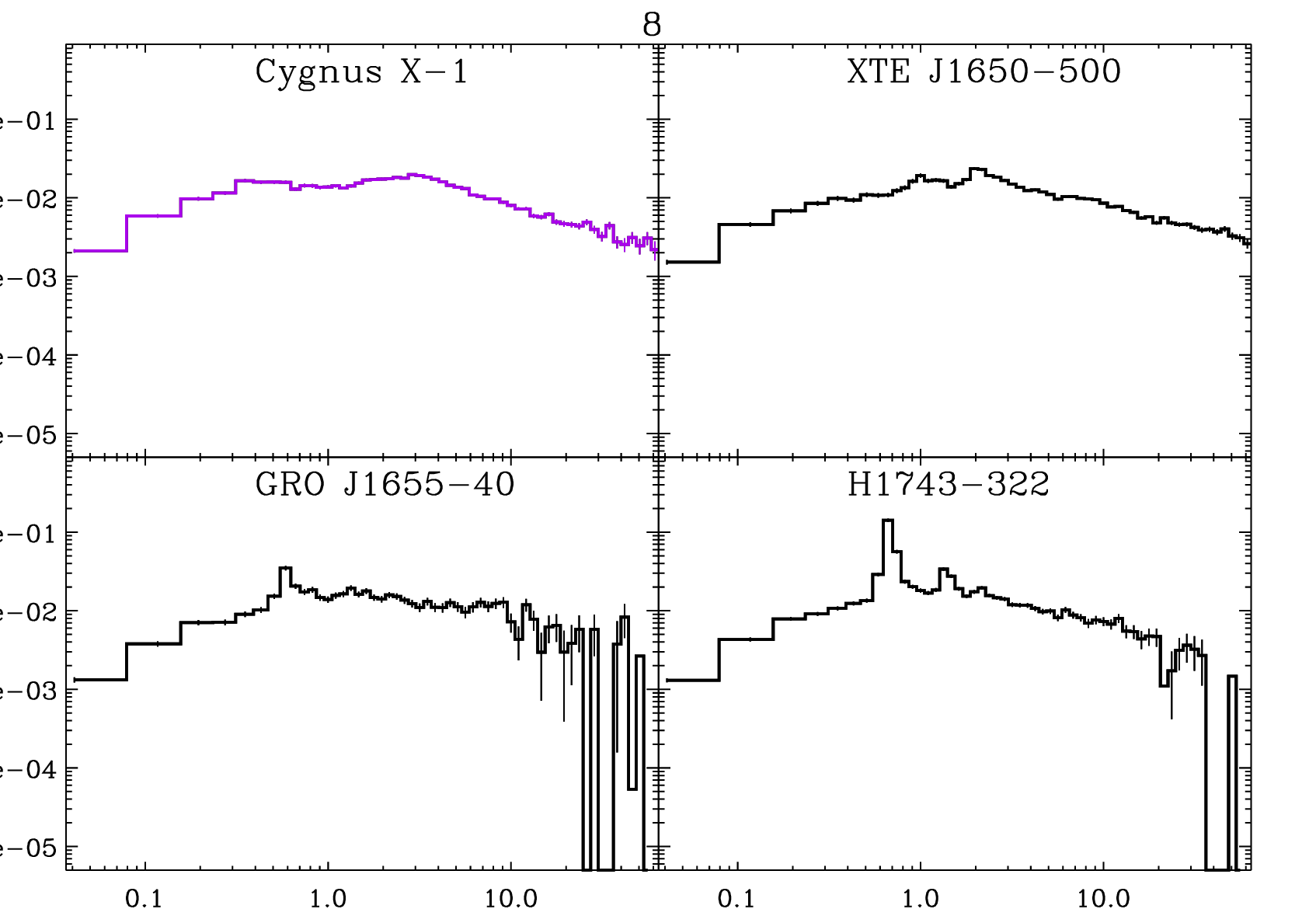}
\end{array}$
\end{center}
\end{figure*}

\begin{figure*}
\begin{center}$
\begin{array}{ll}
	\includegraphics[width=6.0 cm, angle=90]{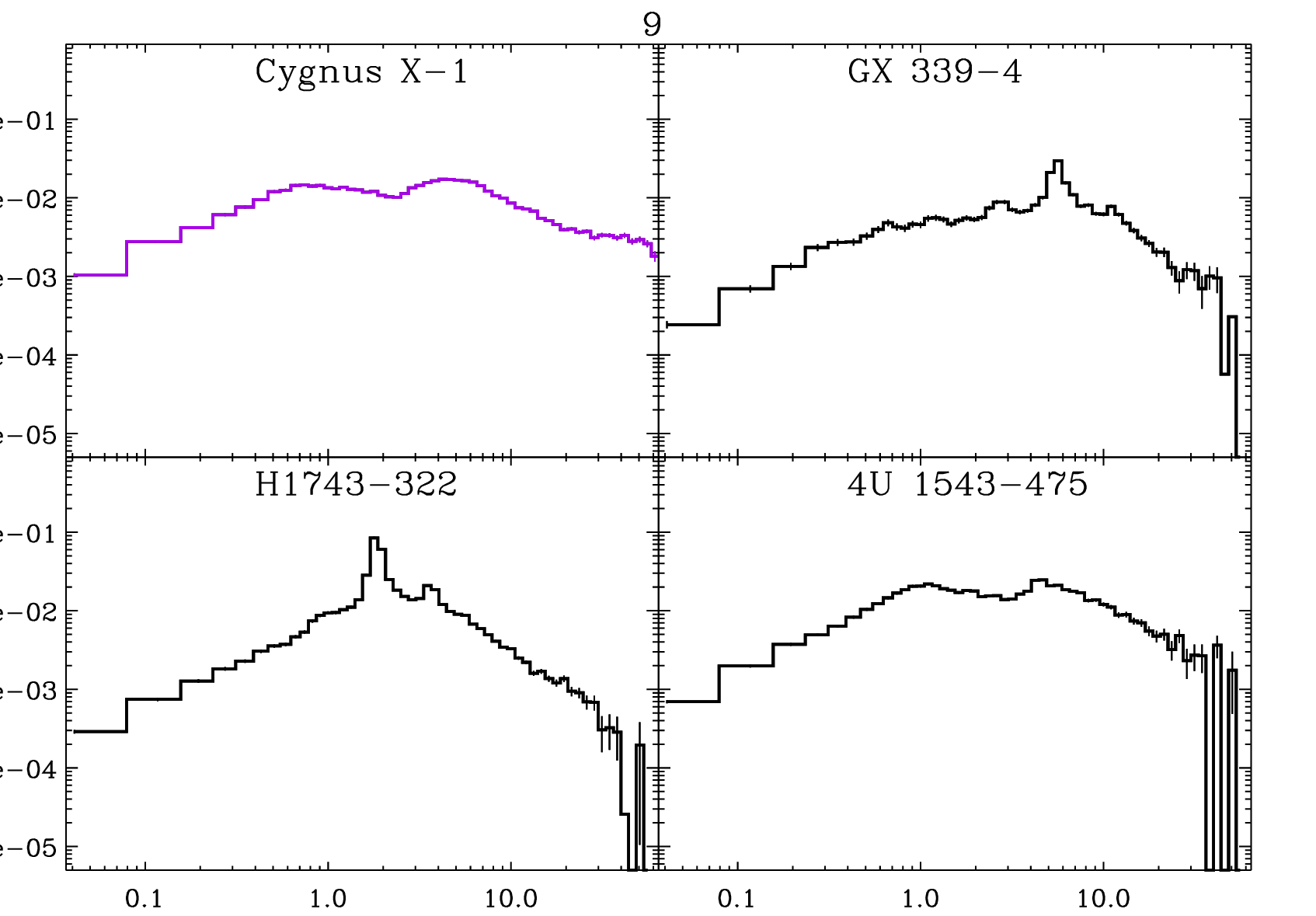} &
	\includegraphics[width=6.0 cm, angle=90]{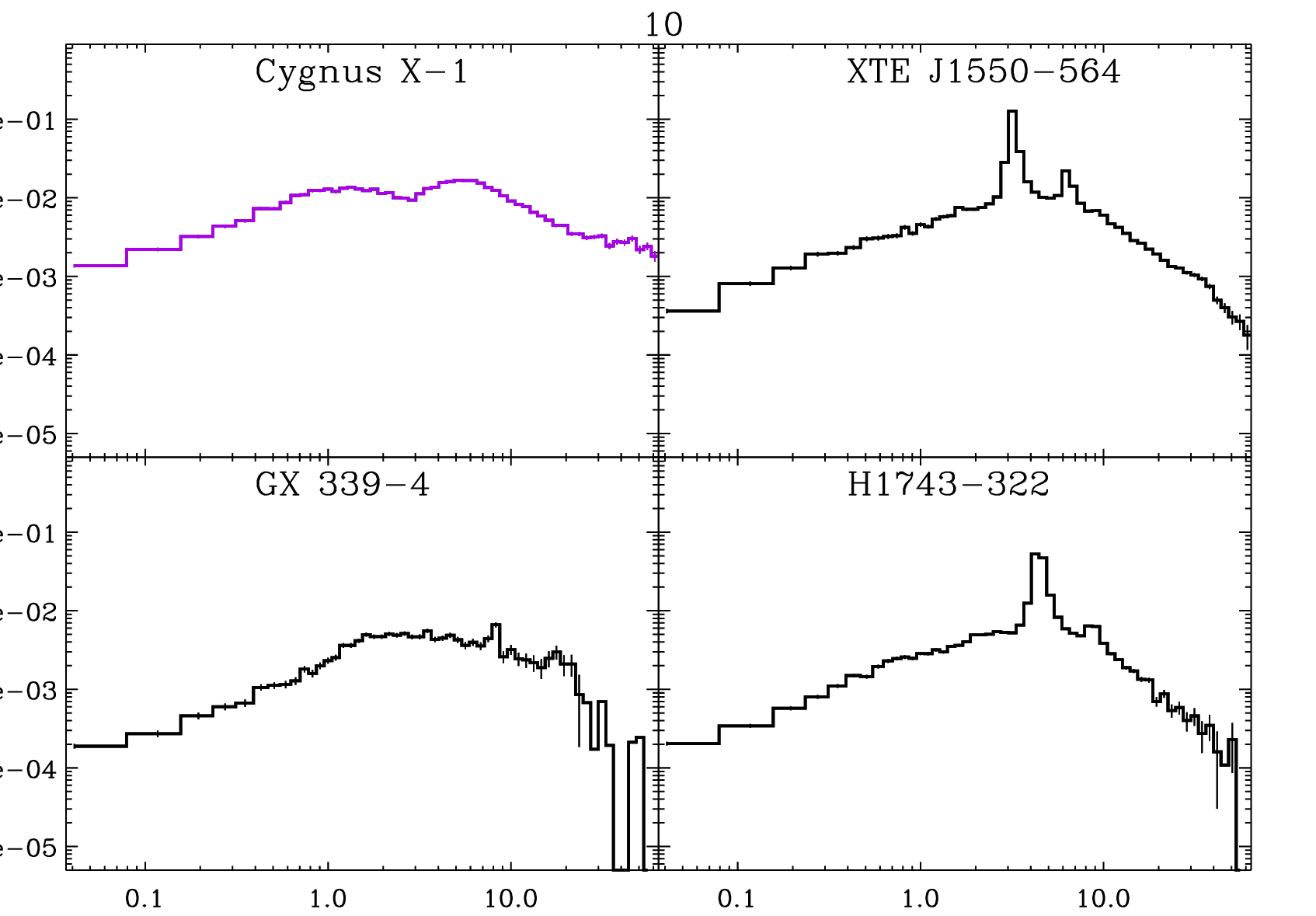}
\end{array}$
\end{center}
\end{figure*}

\begin{figure*}
\begin{center}$
\begin{array}{ll}
	\includegraphics[width=6.0 cm, angle=90]{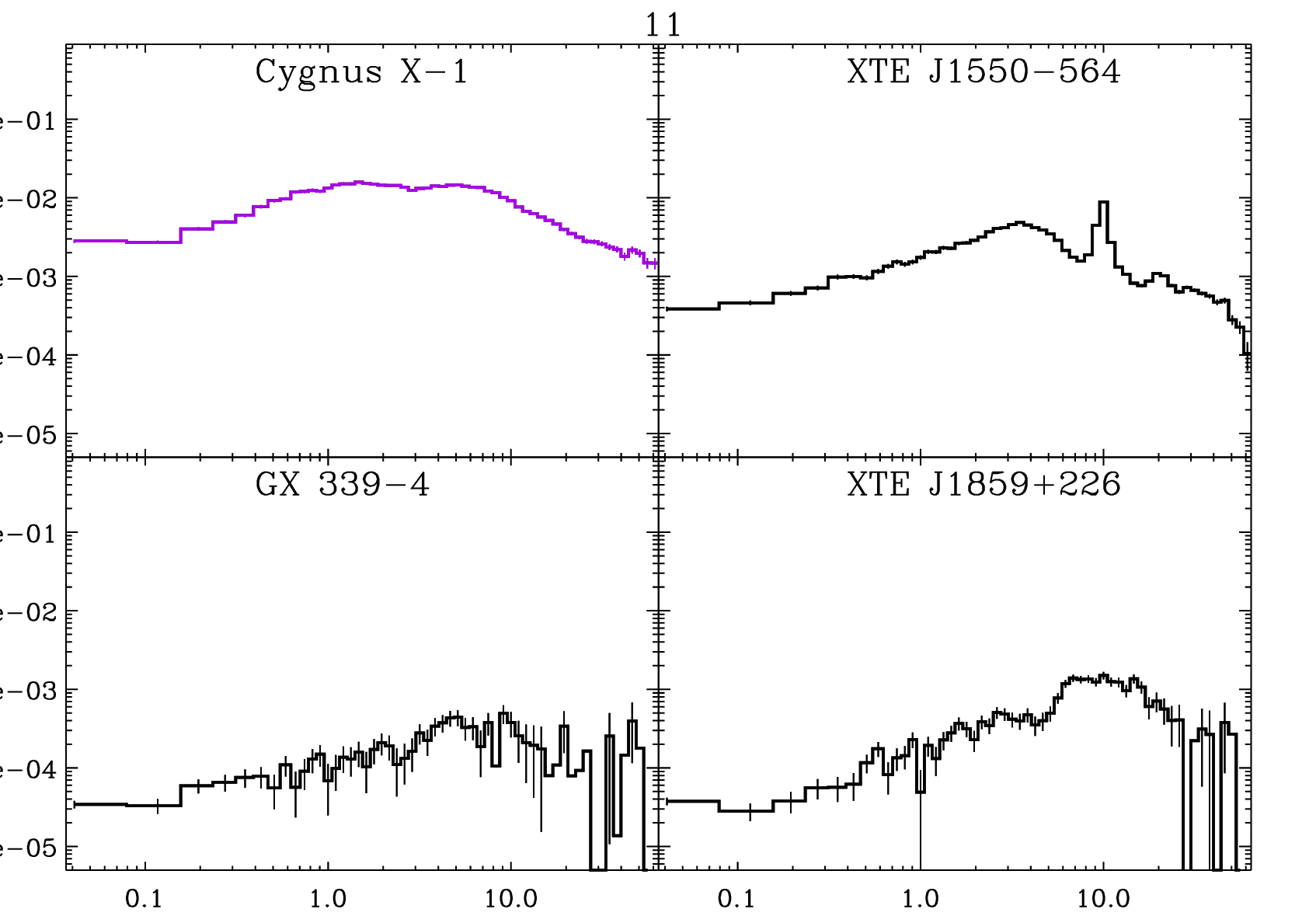} &
	\includegraphics[width=6.0 cm, angle=90]{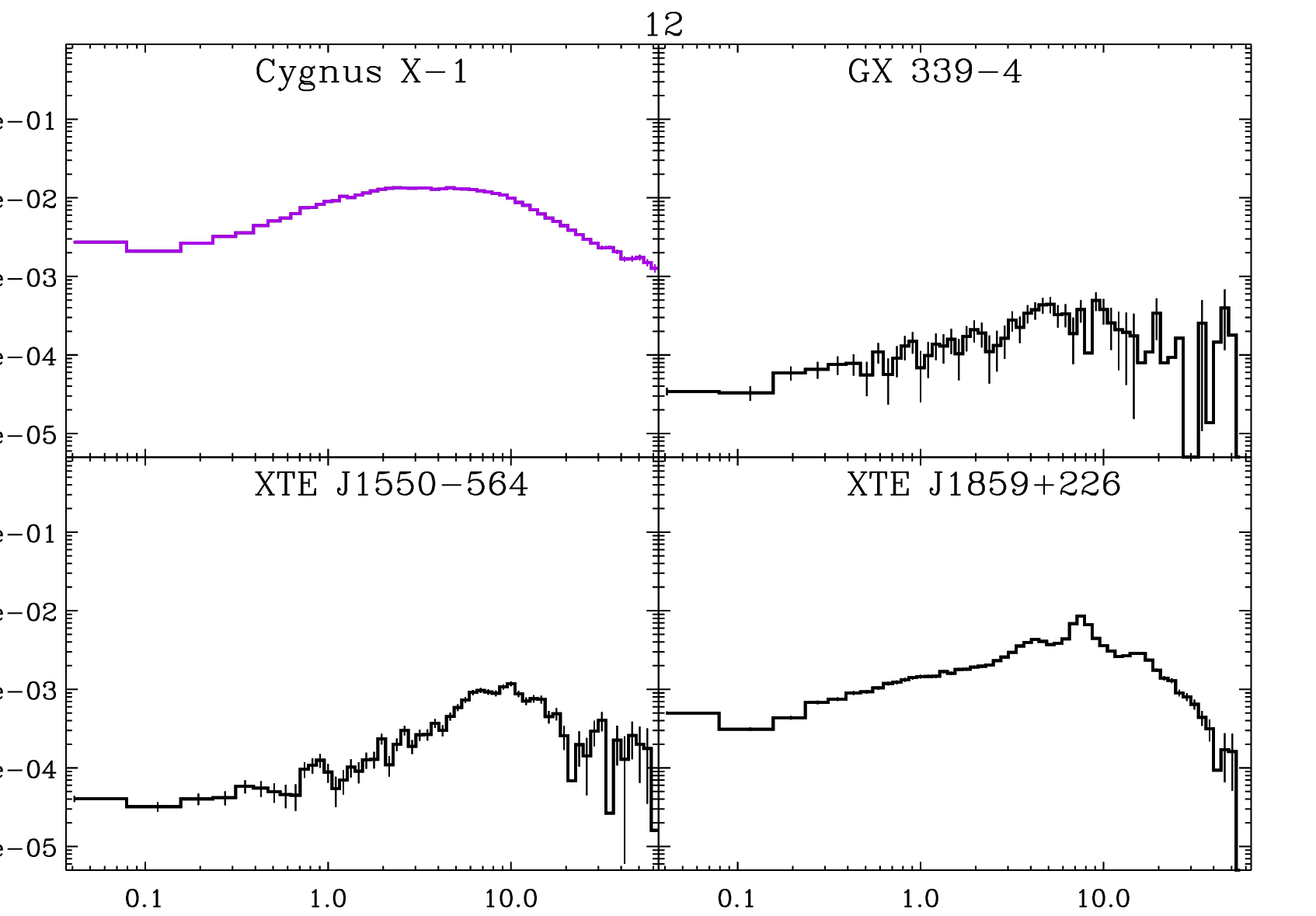}
\end{array}$
\end{center}
\end{figure*}

\begin{figure*}
\begin{center}$
\begin{array}{ll}
	\includegraphics[width=6.0 cm, angle=90]{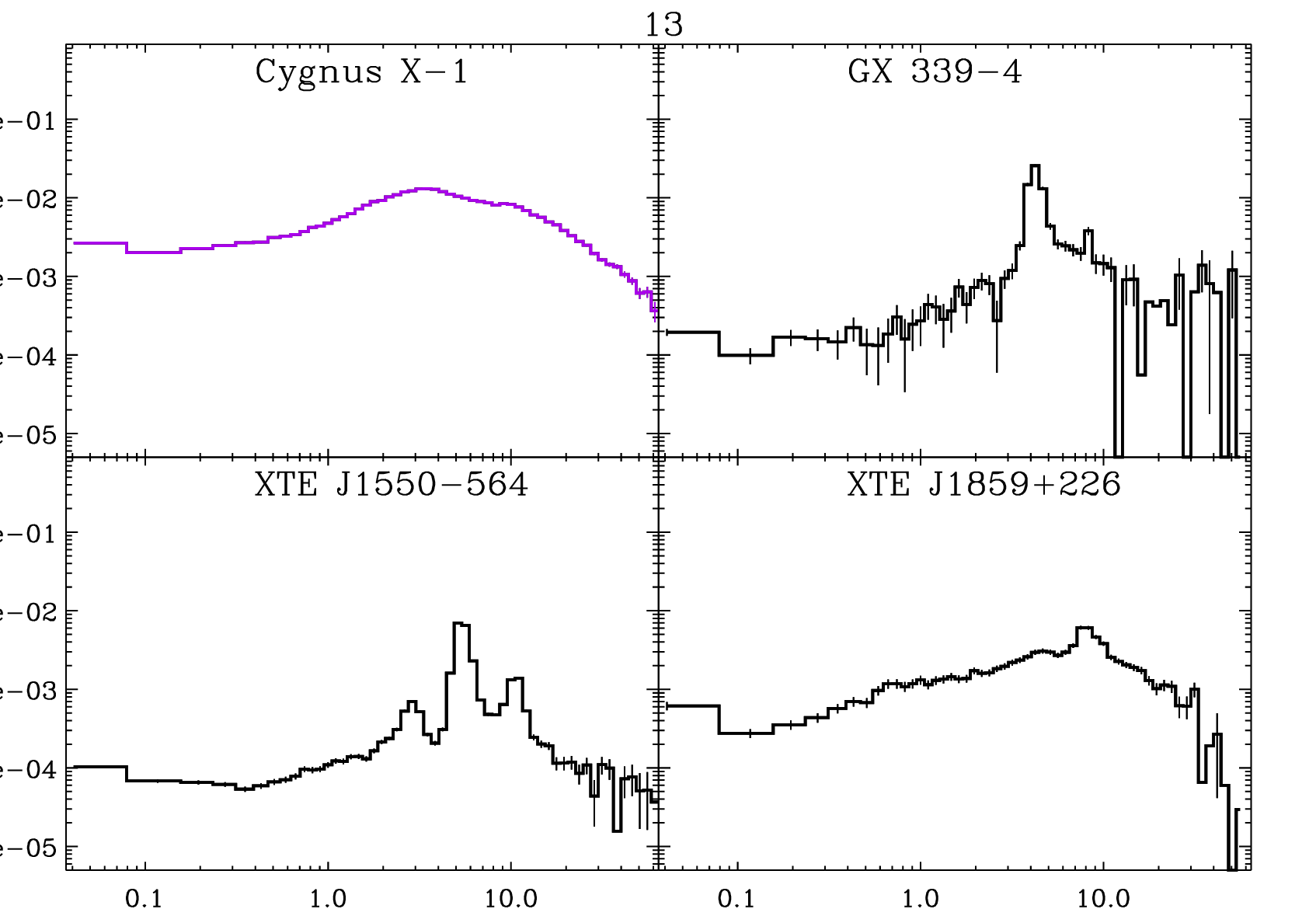} &
	\includegraphics[width=6.0 cm, angle=90]{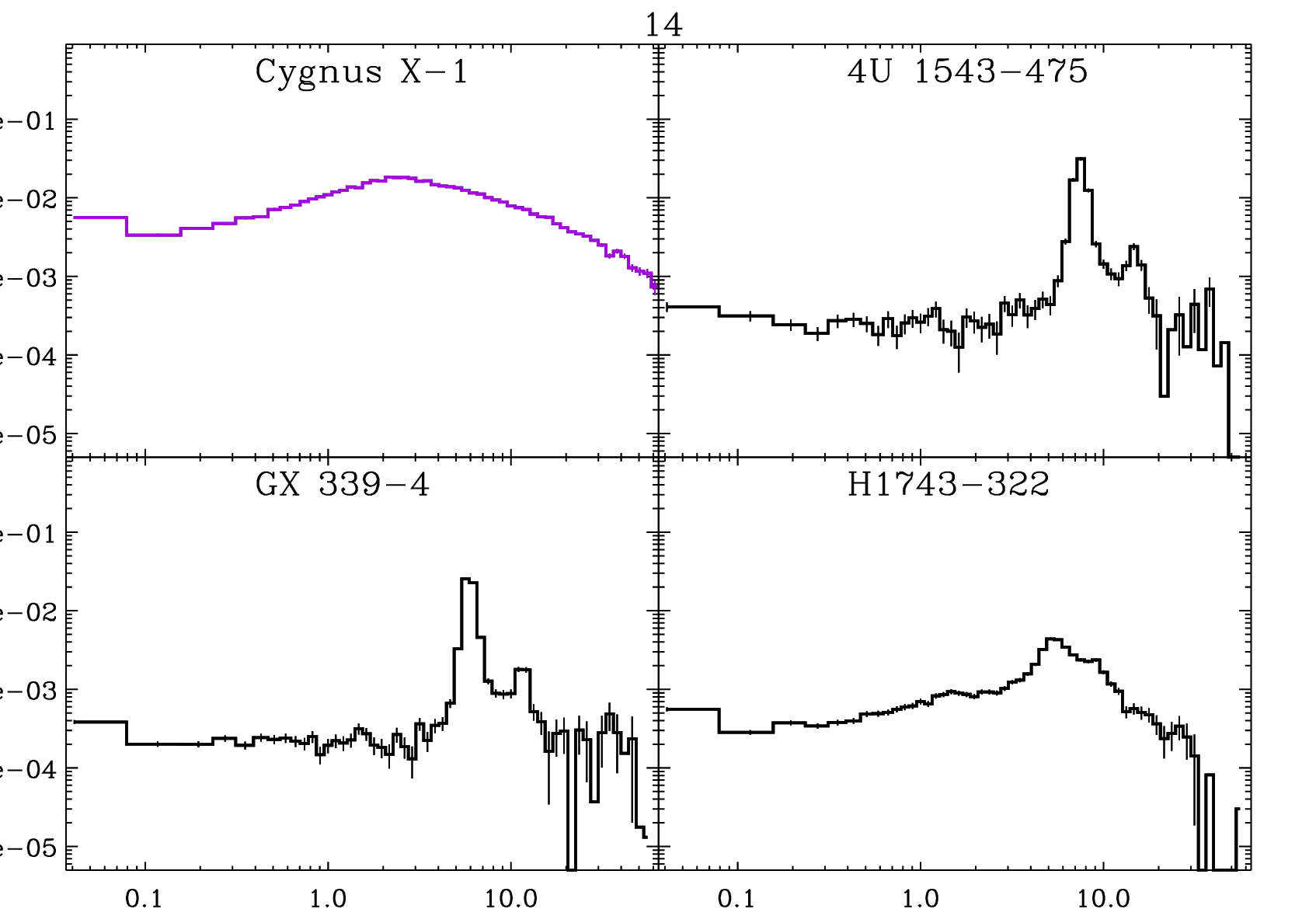}
\end{array}$
\end{center}
\end{figure*}

\begin{figure*}
\begin{center}$
\begin{array}{ll}
	\includegraphics[width=6.0 cm, angle=90]{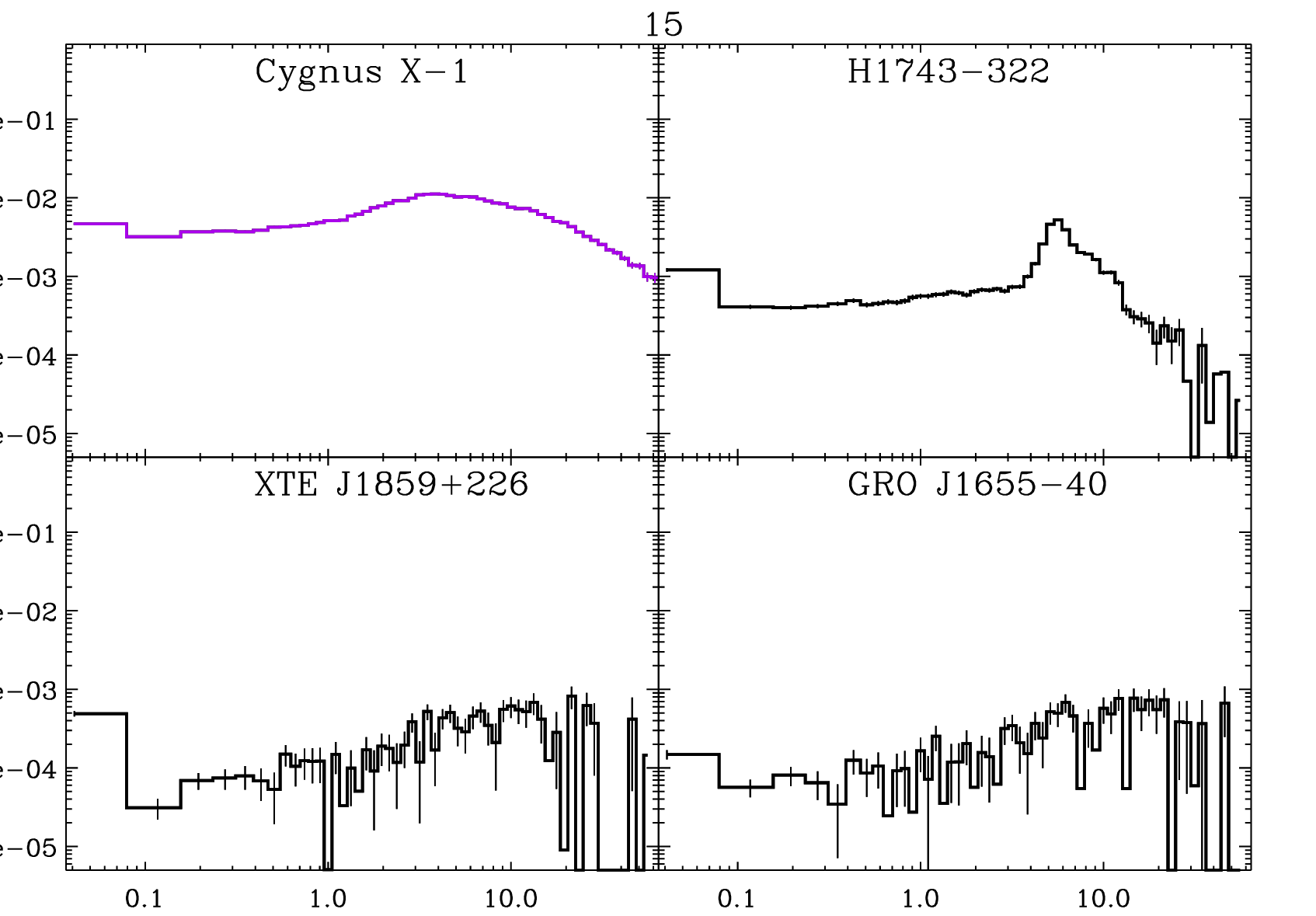} &
	\includegraphics[width=6.0 cm, angle=90]{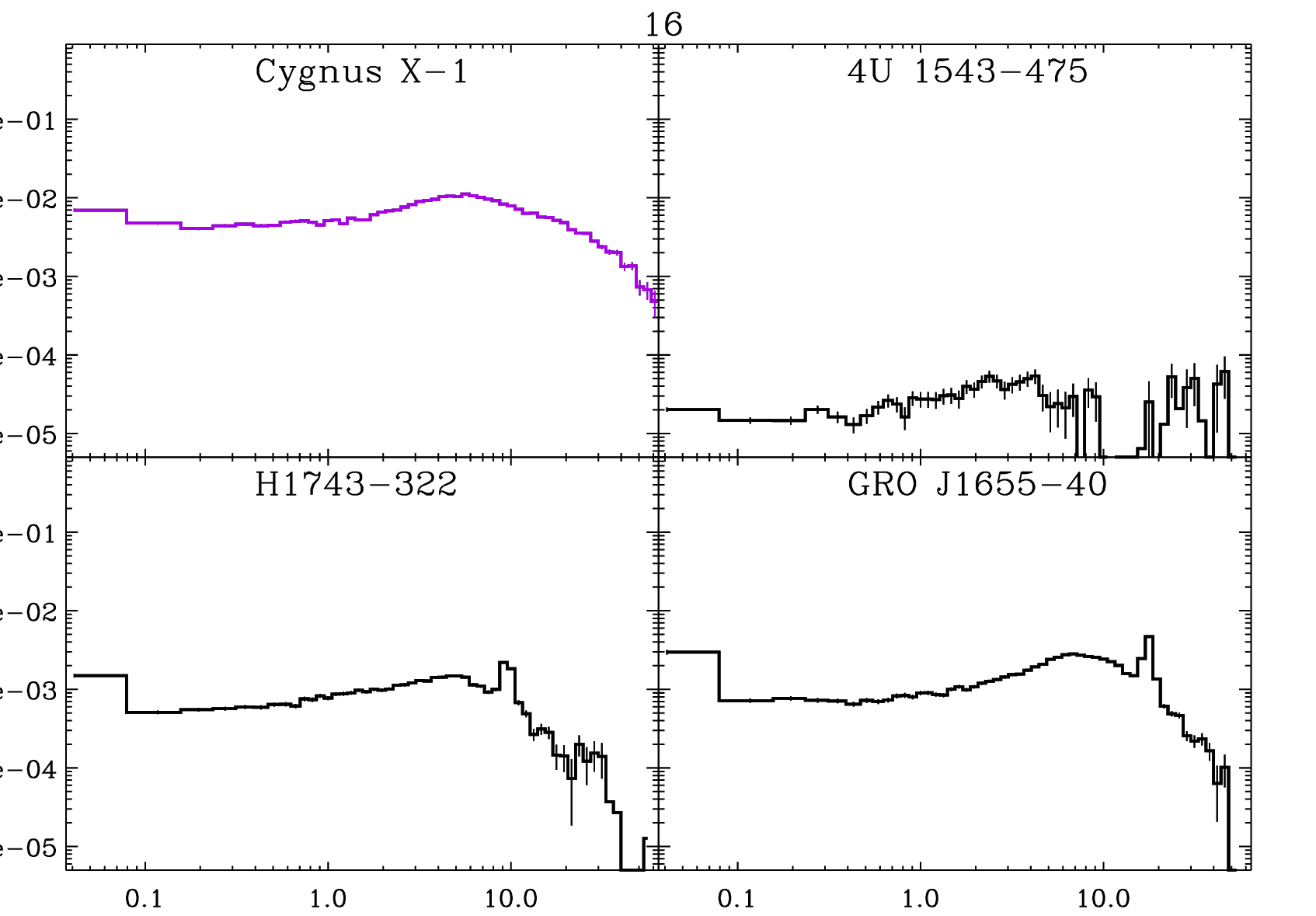}

\end{array}$
\end{center}
\end{figure*}

\begin{figure*}
\begin{center}$
\begin{array}{ll}
	\includegraphics[width=6.0 cm, angle=90]{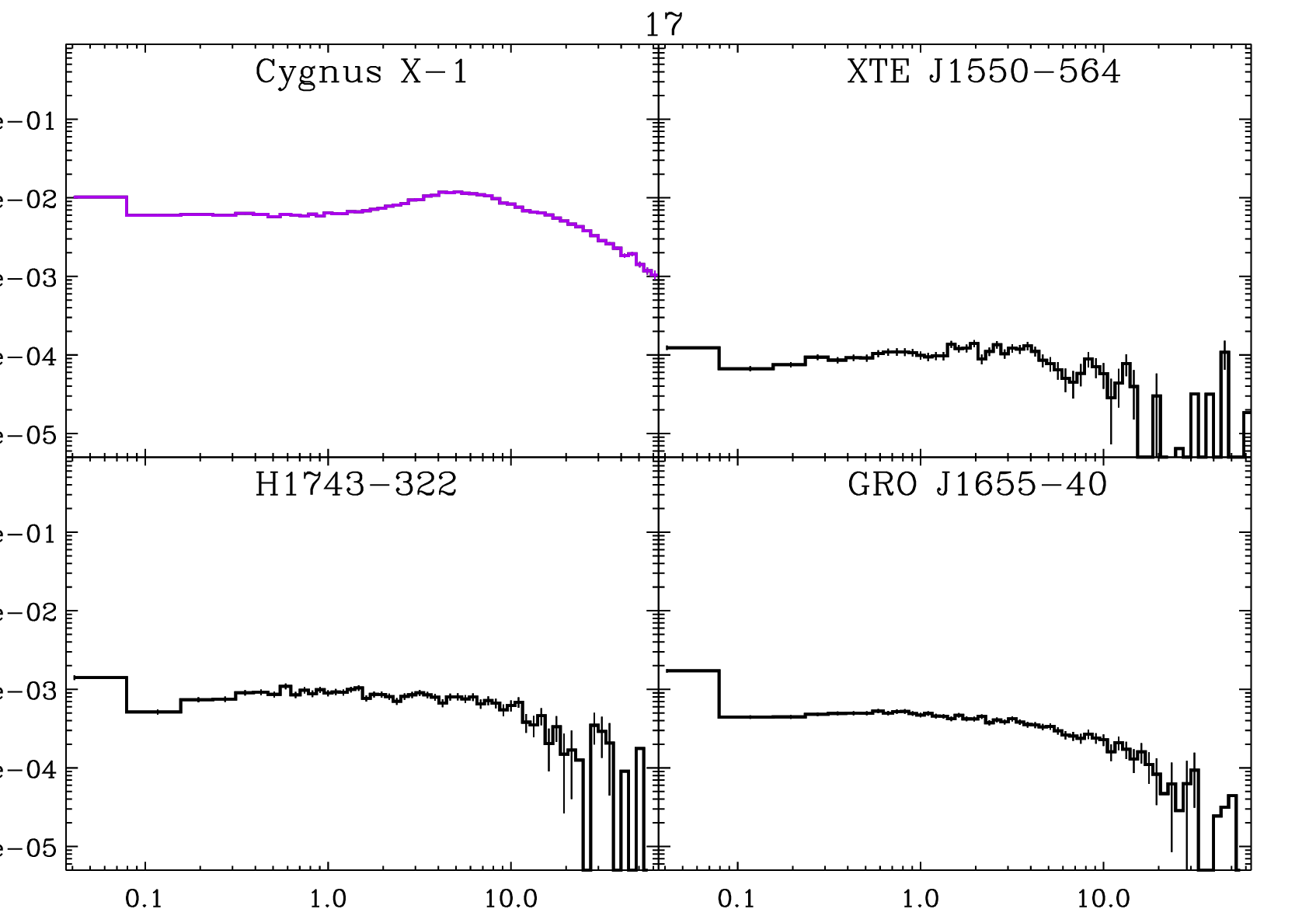} &
	\includegraphics[width=6.0 cm, angle=90]{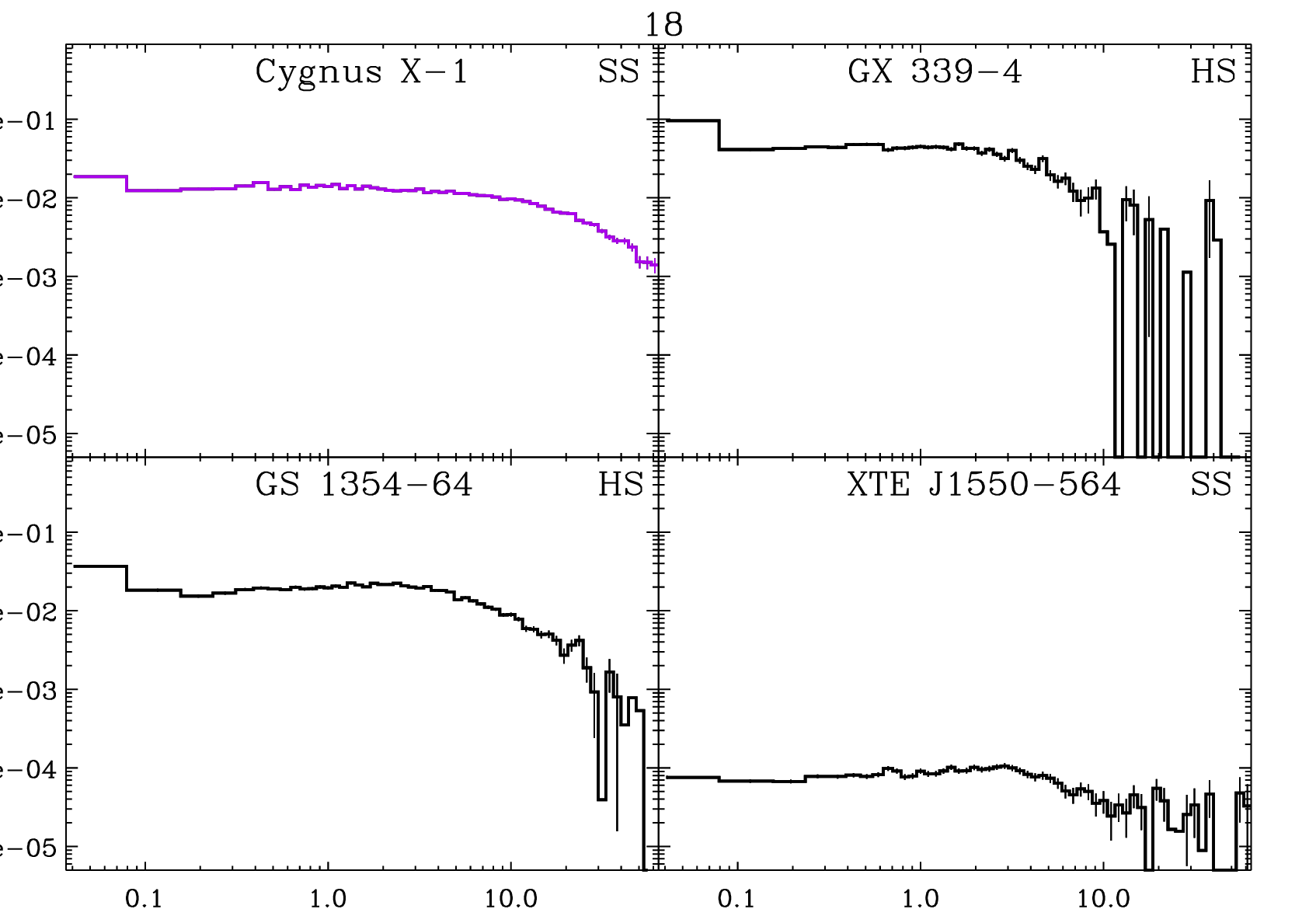}
\end{array}$
\end{center}
\end{figure*}

\label{lastpage}

\end{document}